\documentclass[preprintnumbers,floatfix,aps]{revtex4}
\usepackage{amsmath}
\usepackage{amsfonts}
\usepackage{amssymb}
\usepackage{physymb}
\usepackage{graphicx}
\usepackage{epsfig}
\usepackage{dcolumn}
\usepackage{bm}
\usepackage{fancybox}
\usepackage{multirow}
\usepackage{float}
\usepackage{hyperref}
\usepackage{epstopdf}
\usepackage[utf8]{inputenc}
\usepackage[usenames]{color}

\newcommand{\Punkte}{0}

\newenvironment{Exercise*}[2]{\noindent {\bf Exercise {#1}*.} #2 \vspace{0.2cm} \\
renewcommand{\Punkte}{#2}}{\mbox{\hspace{2ex}} \hfill {\bf \Punkte~\mbox{Points}}\bigskip }

\newcounter{enum1}

\newcounter{enuma}

\begin{document}

\title{\textbf{Solitons and Vortices in Two-dimensional Discrete Nonlinear
Schr{\"{o}}dinger Systems with Spatially Modulated Nonlinearity}}
\date{\today}
\author{P. G. Kevrekidis \thanks{%
Email: kevrekid@math.umass.edu}}
\affiliation{Department of Mathematics and Statistics, University of Massachusetts,
Amherst, MA 01003-4515, USA}
\affiliation{Center for Nonlinear Studies and Theoretical Division, Los Alamos National
Laboratory, Los Alamos, NM 87544}
\author{Boris A. Malomed}
\affiliation{Department of Physical Electronics, School of Electrical Engineering,
Faculty of Engineering, Tel Aviv University, Tel Aviv 69978, Israel}
\author{Avadh Saxena}
\affiliation{Center for Nonlinear Studies and Theoretical Division, Los Alamos National
Laboratory, Los Alamos, New Mexico 87545, USA}
\author{D. J. Frantzeskakis}
\affiliation{Department of Physics, University of Athens, Panepistimiopolis, Zografos,
Athens 15784, Greece}
\author{A.R. Bishop}
\affiliation{Center for Nonlinear Studies and Theoretical Division, Los Alamos National
Laboratory, Los Alamos, New Mexico 87545, USA}

\begin{abstract}
We consider a two-dimensional (2D) generalization of a recently proposed model 
[Phys. Rev. E \textbf{88}, 032905 (2013)], which gives rise to
bright discrete solitons supported by the defocusing nonlinearity whose
local strength grows from the center to the periphery. We explore the 2D model
starting from the anti-continuum (AC) limit of vanishing coupling. In this
limit, we can construct a wide variety of solutions including not only
single-site excitations, but also dipole and quadrupole ones.
Additionally, two separate families of solutions are explored: the usual
``extended'' unstaggered bright solitons,
in which all sites are excited in the AC limit, with the same sign across
the lattice (they represent the most robust states supported by the lattice,
their 1D counterparts being what was considered as 1D bright solitons in the
above-mentioned work), and the vortex cross, which is specific to the 2D
setting. For all the existing states, we explore their stability
(analytically, whenever possible). Typical scenarios of instability
development are exhibited through direct simulations.
\end{abstract}

\maketitle


\section{Introduction}

In the past few years, a topic that has drawn an ever-increasing amount of
interest in the realm of physical systems modeled by nonlinear-Schr{\"{o}}%
dinger (NLS) type equations concerns the examination of solitary waves and
their existence, stability and dynamical properties in the presence of
spatially inhomogeneous nonlinearities. A review which covers many aspects
of this topic can be found in~Ref. \cite{borisreview}. A ramification that
is gaining increased attention within this broader theme concerns the
possibility of the existence of bright coherent structures in the context of
\emph{defocusing} nonlinearities. As is well-known~\cite{ablowitz,kivshar,emergent}, 
systems with a self-defocusing nonlinearity support wave excitations in
the form of dark solitons, vortices, vortex rings etc., i.e., structures
supported by a non-vanishing background at infinity. However, a fundamental
proposal put forth a few years ago 
\cite{malom1,malom2,malom3,Dmitry}, was that, if the local strength of the
self-defocusing in the $D$-dimensional space grows with distance $r$ from
the center at any rate faster than $r^{D}$, then bright solitary waves and
vortical structures can self-trap within such settings. Subsequently, this
class of models was extended to include spatially inhomogeneous nonlinear
losses~\cite{malom4}, higher-power (e.g., quintic) nonlinearities~\cite{malom5}, 
other waveforms such as domain walls~\cite{malom6}, as well as
settings related to Fermi and Bose gases~\cite{malom7}, dipolar
Bose-Einstein condensates~(BECs) \cite{malom8}, nonlocal media~\cite{malom9},
discrete systems~\cite{malom10}, and complex three-dimensional (3D)
topological patterns~\cite{malom11}. Most recently, the Bose-Hubbard model
with the same type of spatial modulation of the self-repulsive
nonlinearity was introduced, and existence of the respective quantum
discrete solitons was demonstrated in it \cite{Padova}.

Another area which has drawn major interest over the past two decades is the
study of models based on the discrete NLS (DNLS) equation~\cite{book}. DNLS
systems have been serving not only as fundamental dispersive systems 
combining nonlinearity and discreteness, but also as models suitable for
the direct description of dynamics in arrays of
 optical waveguides~\cite{dnc,moti} 
and atomic BECs loaded into optical lattices~\cite{ober}. There
are numerous other applications of DNLS models, ranging from 
their use as envelope equations for
understanding the denaturation of the DNA double strand~\cite{Peybi}, 
and the 
localization of energy in granular crystals~\cite{theo10,darkbreath}, to
the dynamics of protein loops~\cite{niemi}.

Our aim in the present work is to combine these two important directions by
extending the 1D model and analysis presented in a recent work \cite{malom10}
to 2D lattices. We will also develop a different approach, examining the
problem from the perspective of the well-established anti-continuum (AC)
limit 
\cite{mackay}, which offers two important advantages. On the one
hand, in the AC limit, which corresponds to vanishing coupling between the
nearest neighbors, we are able to construct solutions systematically, by
initially exciting a single site, multiple sites (two for dipole
configurations, or four for quadrupole ones), as well as possibly all sites in
what we refer to as an extended solution. The same approach allows to
produce not only real waveforms (with relative phases $0$ or $\pi $ between
adjacent sites), but also complex ones, such as discrete vortices. The latter, 
have not only been theoretically proposed~\cite{johan,malomdv}, but
also experimentally observed in photorefractive crystals as per the
theoretical prediction~\cite{neshev,fleischer}. The second important
advantage is that, following the methodology of~Refs.~\cite{pkf,pkf2}, we
are able to provide a systematic classification of the spectral stability of
the states, while departing from the AC limit. In this way, we are able to
predict which states are robust near this limit. We also numerically
corroborate these predictions and, finally, we use direct simulations to
explore the outcome of the evolution of unstable states.

The presentation of the paper is structured as follows. In section II, we introduce the
model and present the theoretical analysis of the existence and stability of
different states. In section III, we explore the model in terms of the
numerically implemented bifurcation theory (as concerns the existence and
spectral stability), and report results of direct simulations of unstable
states. In Section IV, we summarize our findings and discuss directions for
future research.

\section{The Model and Its Analysis}

Generalizing to 2D the considerations of~Ref. \cite{malom10}, we consider a DNLS 
model of 
the following general form:
\begin{equation}
i\dot{u}_{m,n}=-\varepsilon \left(
u_{m,n-1}+u_{m,n+1}+u_{m+1,n}+u_{m-1,n}-4u_{m,n}\right)
+g(m,n)|u_{m,n}|^{2}u_{m,n},  \label{d2dnls1}
\end{equation}
where $\varepsilon $ accounts for the coupling between adjacent wells, and $%
g(m,n)$ represents the local strength of the nonlinearity. Prototypical
examples represent arrays of waveguides in LiNbO$_{3}$~\cite{kip1,kip2,kip3}
and atomic BECs (e.g., of 
$^{87}$Rb or $^{23}$Na gases) confined in an optical lattice in the
superfluid regime~\cite{trosme,alfim}. As argued in~Ref. \cite{borisreview},
a local modulation of the Kerr coefficient in optics, or  a spatial
modulation of the scattering length in atomic BECs (via the Feshbach
resonance) straightforwardly leads to settings of the type we consider here.

In the present section, we develop the analysis in the general form. For the
numerical investigation of section III, we resort to a specific form of the
spatial modulation,
\begin{equation}
g(m,n)=\exp \left( 2(|m|+|n|)\right) ,  \label{g}
\end{equation}
which is a counterpart of the 1D modulation format adopted in 
Ref.~\cite{malom10}. We will also often compare our findings to those in the
homogeneous lattice with $g(m,n)=1$, where solely \textit{staggered}
solitary modes \cite{book} can be obtained for the presently considered
nonlinearity of the defocusing sign.

\begin{figure}[th]
\par
\begin{center}
\includegraphics[height=6cm,width=0.32\textwidth]{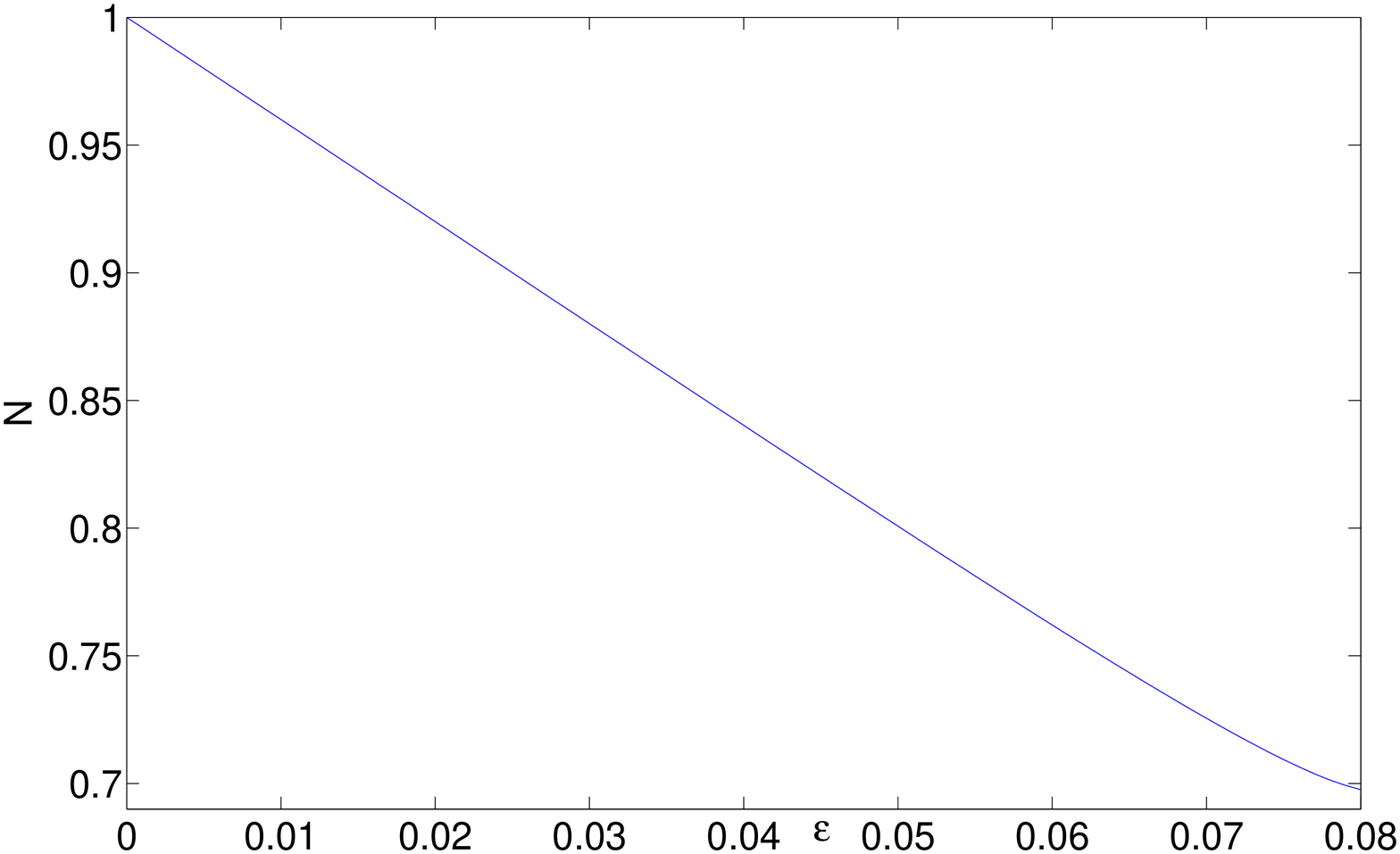} %
\includegraphics[height=6cm,width=0.32\textwidth]{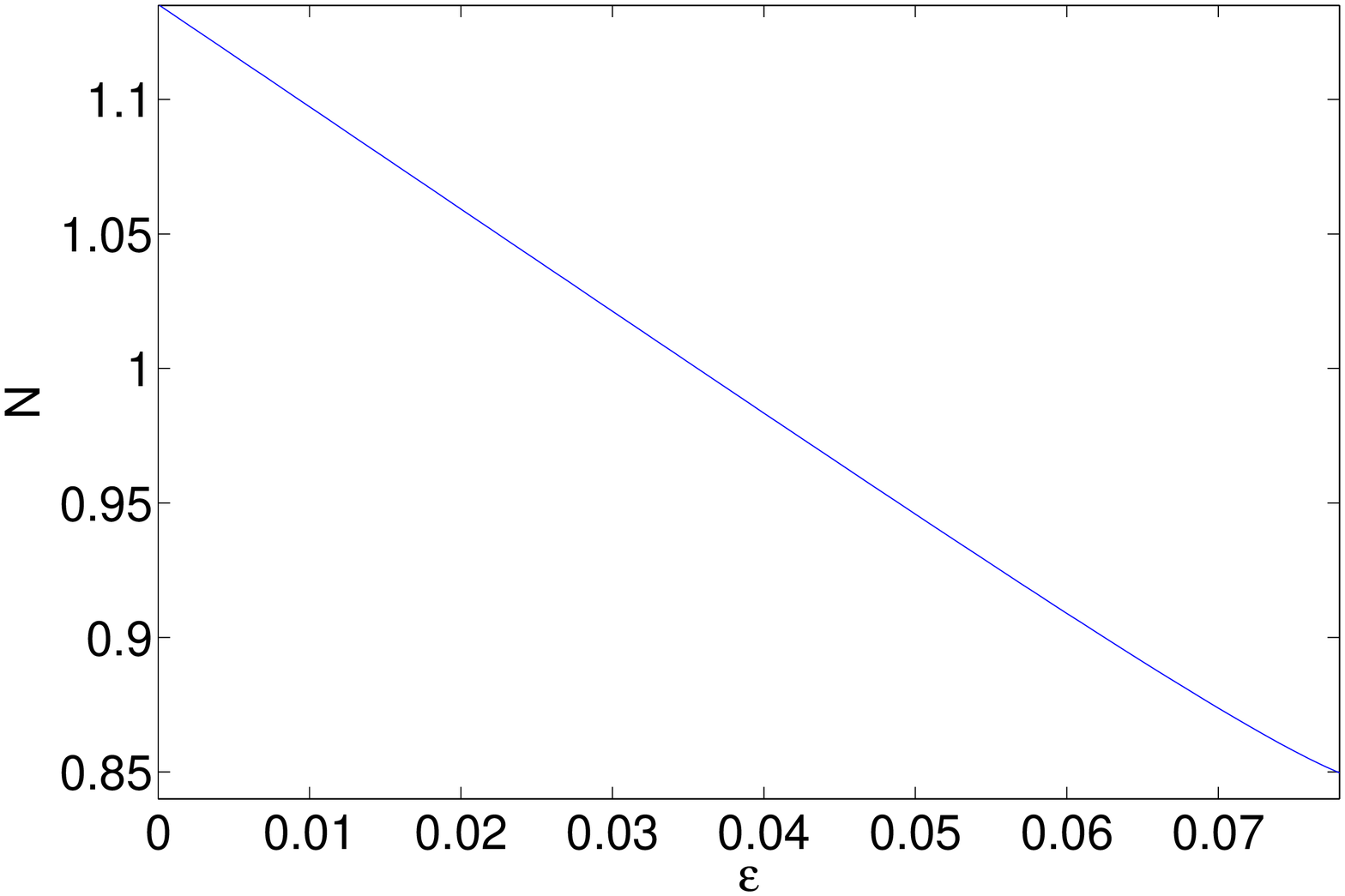} %
\includegraphics[height=6cm,width=0.32\textwidth]{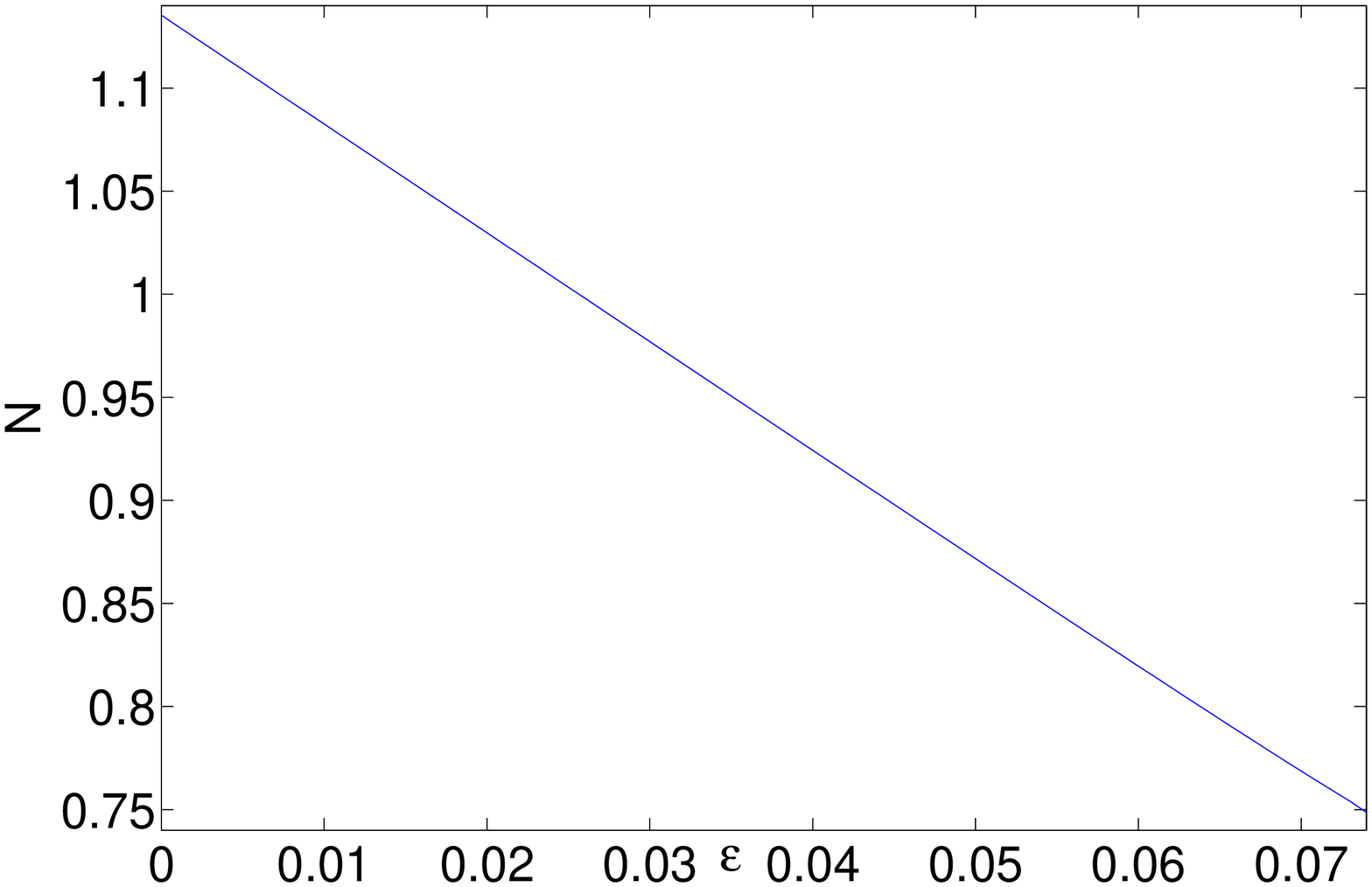}
\end{center}
\caption{(Color Online) 
Numerically generated dependences $N(\protect\epsilon )$ for the
families of single-site, in-phase two-site, and out-of-phase two-site modes
are shown in the left, middle, and right panels, respectively.}
\label{new_fig1a}
\end{figure}

Our first aim is to construct stationary states in the form of $%
u_{m,n}=e^{-i\mu t}v_{m,n}$ with chemical potential (in terms of the BEC) $\mu >0
$, which leads to an equation for $v_{m,n}$:
\begin{equation}
\mu v_{m,n}=-\varepsilon \Delta _{2}v_{m,n}+g(m,n)|v_{m,n}|^{2}v_{m,n},
\label{d2dnls2}
\end{equation}
with $\Delta _{2}v_{m,n}\equiv
v_{m,n-1}+v_{m,n+1}+v_{m+1,n}+v_{m-1,n}-4v_{m,n}$. The total norm of the
mode is defined in the usual form,
\begin{equation}
N=\sum_{m,n}|u_{m,n}|^{2},  \label{N}
\end{equation}
and is a conserved quantity of the model. Families of stationary
solutions are characterized below by dependences $N(\varepsilon )$
for $\mu \equiv 1$, see Fig.~\ref{new_fig1a}. It is also possible
to cast these dependences into the form of $N(\mu )$ for 
$\varepsilon \equiv 1$: as follows from Eqs.~(\ref{d2dnls2}) and 
(\ref{N}), obvious rescaling yields 
\begin{equation}
N(\varepsilon,\mu)=\varepsilon N(1,\mu/\varepsilon) = \mu N(\varepsilon/\mu,1).
\label{NN}
\end{equation}
Applying this for $\varepsilon=1$, we obtain $N(1,\mu)=\mu N(1/\mu,1)$.
In this connection, it is relevant to mention that a necessary
stability condition for solitary modes supported by repulsive nonlinearities
is given by the anti-Vakhitov-Kolokolov criterion \cite{anti}, $dN/d\mu >0$. 
In particular, nearly linear dependences of $N$ on $\varepsilon$ 
observed in Fig.~\ref{new_fig1a}, if substituted into Eq. (\ref{NN}),
correspond to $dN/d\mu \approx N(\varepsilon =0,\mu)>0$. Indeed, actual
results for the stability reported below confirm that the particular
instability mechanism, which may be detected by the anti-Vakhitov-Kolokolov
criterion, is absent in the present system. 

The starting point of the analysis is the AC limit of $\varepsilon =0$, corresponding to the case 
where the sites get decoupled. In this limit, the only local solutions corresponding -- respectively -- 
to non-excited or excited sites, 
are: $v_{m,n}=0$ or
\begin{equation}
v_{m,n}=\sqrt{\mu /g(m,n)}e^{i\theta _{m,n}}.  \label{TF}
\end{equation}
%
Equation~(\ref{TF}) yields, in fact, the Thomas-Fermi
approximation (TFA) for the lattice field \cite{malom10,Dmitry}, which, in
particular, leads to the conclusion that the solution is normalizable (in
other words, it is a 
physically relevant one), i.e., its norm (\ref{N}) converges,
under the condition that $g\left( m,n\right) $ must grow, as $%
|m|,|n|\rightarrow \infty $, at any rate faster than $\left(
m^{2}+n^{2}\right)$.

Based on this AC-limit solution, we can choose to excite any configuration
in the AC limit, with an arbitrary phase pattern. The actual issue is which
ones of these configurations persist at finite values of inter-site coupling
$\varepsilon $. To address it, works \cite{pkf}, \cite{pkf2}, and~\cite{pkf3}
(for 1D, 2D, and 3D cubic lattices, respectively) have developed a
``persistence condition", which we now adapt to the present setting.

Suppose that a string of three sites is excited, with coordinates 
$(m,n-1)$, $(m,n)$ and $(m,n+1)$. Then the persistence condition, 
adapted to the present setting, reads:
\begin{equation}
\frac{\sin (\theta _{m,n}-\theta _{m,n-1})}{\sqrt{g(m,n)g(m,n-1)}}=\frac{%
\sin (\theta _{m,n+1}-\theta _{m,n})}{\sqrt{g(m,n+1)g(m,n)}}.
\label{d2dnls3}
\end{equation}
Pertaining to two-point functions defined for adjacent pairs of sites, it
can be generalized for any set of such pairs of sites.

In 1D, given that
 this set of two-point functions is the same for all sites up to
$\pm \infty $, for solutions that vanish at infinity, the persistence
condition allows only configurations with relative phases $0$ or $\pi $.
However, this is no longer the case in 2D, as the condition can be satisfied
over closed contours without the need to extend the considerations to
infinity. As a result, in the latter setting complex configurations,
including vortices, are possible. Nevertheless, the simpler configurations
are the ones with relative phases $\Delta \theta _{n}=0$ or $\pi $, which we
predominantly consider below.

Particular configurations that we aim to study are:

\begin{enumerate}
\item A single-site solution with $v_{0,0}=\sqrt{\mu /g(0,0)}$ and vanishing
amplitude at all other sites.

\item A ``dipolar" state resting on a pair
of sites, e.g., $(0,0)$ and $(1,0)$. These two sites may be excited 
\textit{in}- or \textit{out}-of-phase.

\item ``Quadrupole" configurations
supported by four sites. Although additional structures can also be
considered (which is true for the dipolar modes too), we restrict our
considerations here to the square-shaped set of four sites $(0,0)$, $(1,0)$,
$(1,1)$, $(0,1)$. Qualitative conclusions that we will infer for the
stability will not be different if we choose another quartet of sites,
although details may differ.

\item We also consider the extended unstaggered solution in which \textit{all%
} sites of the lattice are excited with the same sign, as $v_{m,n}=\sqrt{\mu
/g(m,n)}$, cf. Eq. (\ref{TF}). Actually, the 1D counterpart of such a state
was the subject of~the analysis in Ref. \cite{malom10}, while solutions
which amount to single- or few-site excitations in the AC limit were not
considered in that work.

\item Finally, while for the above-mentioned square-shaped quartet of sites,
$(0,0)$, $(1,0)$, $(1,1)$, $(0,1)$, with $g\left( m,n\right) $ taken even in
both $n$ and $m$, we were unable to continue vortical solutions for finite $%
\varepsilon $, nevertheless, we were able to do so for a cross-shaped
quartet, $(1,0)$, $(0,1)$, $(-1,0)$, $(0,-1)$, which features an empty site
at the center.
\end{enumerate}

Now, we turn to the consideration of the stability of the discrete
configurations. To this end, we employ the usual linearization ansatz for
perturbations with small amplitude $\delta $:
\begin{equation}
u_{m,n}=e^{-i\mu t}\left[ v_{m,n}+\delta e^{\lambda t}
p_{m,n}+ \delta e^{\lambda^{\star} t} q_{m,n}^{\star} \right] ,  \label{d2dnls4}
\end{equation}%
(where $^{\star}$ denotes complex conjugate) 
deriving equations at order $\mathcal{O}(\delta )$ for $(p_{m,n},q_{m,n})$.
For simplicity, we mention here only the ensuing eigenvalue problem in the
case when the unperturbed solution $v_{m,n}$ is real, also using
the decomposition~\cite{pkf} 
$p_{m,n}=a_{m,n}+i b_{m,n}$ and $q_{m,n}=a_{m,n} - i b_{m,n}$
\begin{equation}
\lambda \left(
\begin{array}{c}
a_{m,n} \\
b_{m,n}%
\end{array}%
\right) =\left(
\begin{array}{cc}
0 & \mathcal{L}_{-} \\
-\mathcal{L}_{+} & 0%
\end{array}%
\right) \left(
\begin{array}{c}
a_{m,n} \\
b_{m,n}%
\end{array}%
\right).  \label{d2dnls5}
\end{equation}
In these expressions the linear operators are defined as follows: $\mathcal{L%
}_{-}b_{m,n}=-\varepsilon \Delta _{2}b_{m,n}-\mu
b_{m,n}+g(m,n)v_{m,n}^{2}b_{m,n}$ and $\mathcal{L}_{+}a_{m,n}=-\varepsilon
\Delta _{2}a_{m,n}-\mu a_{m,n}+3g(m,n)v_{m,n}^{2}a_{m,n}$. Rewriting the
above non-self-adjoint eigenvalue problem as a combined fourth-order one, we
obtain
\begin{equation}
\lambda ^{2}b_{m,n}=-\mathcal{L}_{+}\mathcal{L}_{-}b_{m,n}\Rightarrow
\lambda ^{2}\mathcal{L}_{+}^{-1}b_{m,n}=-\mathcal{L}_{-}b_{m,n}.
\label{d2dnls6}
\end{equation}
It is relevant now to point out that near the AC limit of $\varepsilon
\rightarrow 0$, $\mathcal{L}_{+}$ becomes a multiplicative operator with
positive entries, which is obviously invertible. Forming the inner product
of Eq.~(\ref{d2dnls6}) with $b_{m,n}$, we obtain
\begin{equation}
\lambda ^{2}=-\frac{\langle b_{m,n},\mathcal{L}_{-}b_{m,n}\rangle }{\langle
b_{m,n},\mathcal{L}_{+}^{-1}b_{m,n}\rangle },  \label{d2dnls7}
\end{equation}
where $\langle , \rangle$ denotes the standard inner product. Given the
multiplicative nature of $\mathcal{L}_{+}$ in the AC limit, the leading-order
approximation near $\varepsilon =0$ yields $\mathcal{L}_{+}^{-1}\rightarrow
(2\mu )^{-1}$ for excited sites with $v_{m,n}\neq 0$ [and $\mathcal{L}%
_{+}^{-1}\rightarrow -(\mu )^{-1}$ for the non-excited ones]. Thus,
eigenvalues of the above-mentioned real solutions are directly associated
with operator $\mathcal{L}_{-}$, up to the above-mentioned multiplicative
factor $-2\mu $ (henceforth, without loss of generality, we will set 
$\mu =1$).

It is straightforward to see that for all the non-excited sites with $%
v_{m,n}=0$, $\mathcal{L}_{-}=-1$, $\lambda =\pm i$. These eigenvalues will
form, as $\varepsilon $ becomes nonzero, the continuous spectrum which, in
the 2D setting, corresponds to the interval $\pm i [1-8\varepsilon ,1]$. 
On the other
hand, the eigenvalues that may lead to instability (at least for small $%
\varepsilon $) are those stemming from the excited sites for which $\mathcal{%
L}_{-}$ vanishes to the leading order, hence these eigenvalues are $\lambda
=0$ at $\varepsilon =0$. In principle, these eigenvalue pairs may become
real immediately as $\varepsilon $ becomes nonzero. It is then of critical
importance, as regards the stability, to identify eigenvalues of 
the matrix $%
M=\langle b,\mathcal{L}_{-}b\rangle \equiv \varepsilon \mathcal{M}$. Upon
obtaining eigenvalues $\gamma $ of the matrix $\mathcal{M}$, based on the
theory presented in Refs. \cite{pkf,pkf2,pkf3} (see also~\cite{book})
and the above exposition, the
eigenvalues $\lambda $ of the full problem will be given, in view of 
Eq.~(\ref%
{d2dnls7}), by $\lambda =\pm \sqrt{-2\varepsilon \gamma }$. We perform this
calculation below for two- and four-site real excitations. For the
single-site excitation, there is only one pair at $\lambda =0$. Actually,
for all configurations one pair always remains at the origin, due to the
phase/gauge invariance of the model (in the case of the single-site
excitation, it is the sole one, so there is no bifurcation occurring). For
the extended solution, since all sites are excited, the number of pairs of
eigenvalues at the origin is equal to the number of nodes in the lattice,
hence the corresponding matrix $\mathcal{M}$ also 
has the same number of rows and
columns. Finally, for the only genuinely complex configuration considered
here, the computation of matrix $\mathcal{M}$ is considerably more
complicated, as it should be performed at a higher order [$\mathcal{O}%
(\varepsilon ^{2})$, rather than $\mathcal{O}(\varepsilon )$, as the
relevant excited sites are two lattice spacings apart and only couple at $%
\mathcal{O}(\varepsilon ^{2})$]. We do not present details of that
calculation here.

In the case of two-site excitations, the matrix $\mathcal{M}$ can be computed
explicitly (upon calculating the leading order i.e., an $\mathcal{O}%
(\varepsilon )$ correction to the solution) as
\begin{equation}
\mathcal{M}=\left(
\begin{array}{cc}
\sqrt{\frac{g(m,n)}{g(m,n+1)}} & -1 \\
-1 & \sqrt{\frac{g(m,n+1)}{g(m,n)}}%
\end{array}%
\right) \cos (\theta _{m,n+1}-\theta _{m,n}).  \label{d2dnls11}
\end{equation}%
Here, we assume that the two excited sites are $(m,n)$ and $(m,n+1)$. The
eigenvalues are then $\gamma =0$ and $\gamma =c\cos (\theta _{m,n+1}-\theta
_{m,n})$, where
\begin{equation}
c\equiv \sqrt{\frac{g(m,n)}{g(m,n+1)}}+\sqrt{\frac{g(m,n+1)}{g(m,n)}}.
\label{c}
\end{equation}
One of them, as indicated above, remains at the origin, while the other
grows along the real axis for out-of-phase excitations (making these immediately
unstable as $\varepsilon $ becomes nonzero) or along the imaginary axis for
in-phase excitations, which does not lead to immediate destabilization.
In both cases, note that the inequality $c\geq 2$ leads to a growth rate for
these eigenvalues which is larger than that of the homogeneous limit of
constant $g(m,n)=1$. Furthermore, even for the in-phase mode, which is
stable for small $\varepsilon $, as the respective imaginary eigenvalues
grow according to $\lambda = \pm i 
\sqrt{2c\varepsilon }$ [recall $c$ is defined by
Eq. (\ref{c})], they eventually collide with the edge of the above-mentioned
continuous spectrum, at $\pm i (1-8\varepsilon) $, leading to an
oscillatory-instability threshold, $\varepsilon =(1/64)\left( 8+c-\sqrt{%
c^{2}+16c}\right) $. Given the larger growth rate of the imaginary
eigenvalue pair bifurcating from the origin, this instability occurs at
smaller values of $\varepsilon $ in comparison to the homogeneous limit of $%
c=2$.

We now turn to the four-excited-site case, which is considerably more
complicated. Here, the reduced matrix $\mathcal{M}$ is of size $4\times 4$.
Labeling the relative phase factors as $r_{10}=\cos (\theta _{1,0}-\theta
_{0,0})$, $r_{21}=\cos (\theta _{1,1}-\theta _{1,0})$, $r_{32}=\cos (\theta
_{0,1}-\theta _{1,1})$, and $r_{03}=\cos (\theta _{0,0}-\theta _{0,1})$, we
can write the matrix:
\begin{equation}
\mathcal{M}=\left(
\begin{array}{cccc}
\sqrt{\frac{g(0,0)}{g(1,0)}}r_{10}+\sqrt{\frac{g(0,0)}{g(0,1)}}r_{03} &
-r_{10} & 0 & -r_{03} \\
-r_{10} & \sqrt{\frac{g(1,0)}{g(0,0)}}r_{10}+\sqrt{\frac{g(1,0)}{g(1,1)}}%
r_{21} & -r_{21} & 0 \\
0 & -r_{21} & \sqrt{\frac{g(1,1)}{g(1,0)}}r_{21}+\sqrt{\frac{g(1,1)}{g(0,1)}}%
r_{32} & -r_{32} \\
-r_{03} & 0 & -r_{32} & \sqrt{\frac{g(0,1)}{g(1,1)}}r_{32}+\sqrt{\frac{g(0,1)%
}{g(0,0)}}r_{03}%
\end{array}%
\right).  
\label{d2dnls12}
\end{equation}

This matrix has a single zero eigenvalue. Furthermore, if all $r$'s are
positive, then the eigenvalues $\gamma $ are also positive, hence the
eigenvalues of the full problem are imaginary at $\varepsilon >0$. On the
other hand, if one (or more) of the relative phase factors $r$ is (are) 
negative,
then the corresponding number of negative $\gamma $'s emerge, leading to
pairs of real eigenvalues, and hence instability of the configuration. These
features are directly in line with what is known for the
homogeneous defocusing model, see, e.g., Ref.~\cite{hadisus}. They are also
directly the reverse of the focusing nonlinearity case (i.e., the features
corresponding to negative $r$ in one case correspond to those for positive $%
r $ in the other). While, in principle, the eigenvalues of this $4\times 4$
matrix can be obtained in an explicit analytical form, the expressions are
too cumbersome to be useful. Therefore we will now turn to numerical
computations, comparing the results with those of the above analysis,
whenever possible.

\section{Numerical Results}

\subsection{Stationary modes and their stability}

\begin{figure}[th]
\par
\begin{center}
\includegraphics[width=0.45\textwidth]{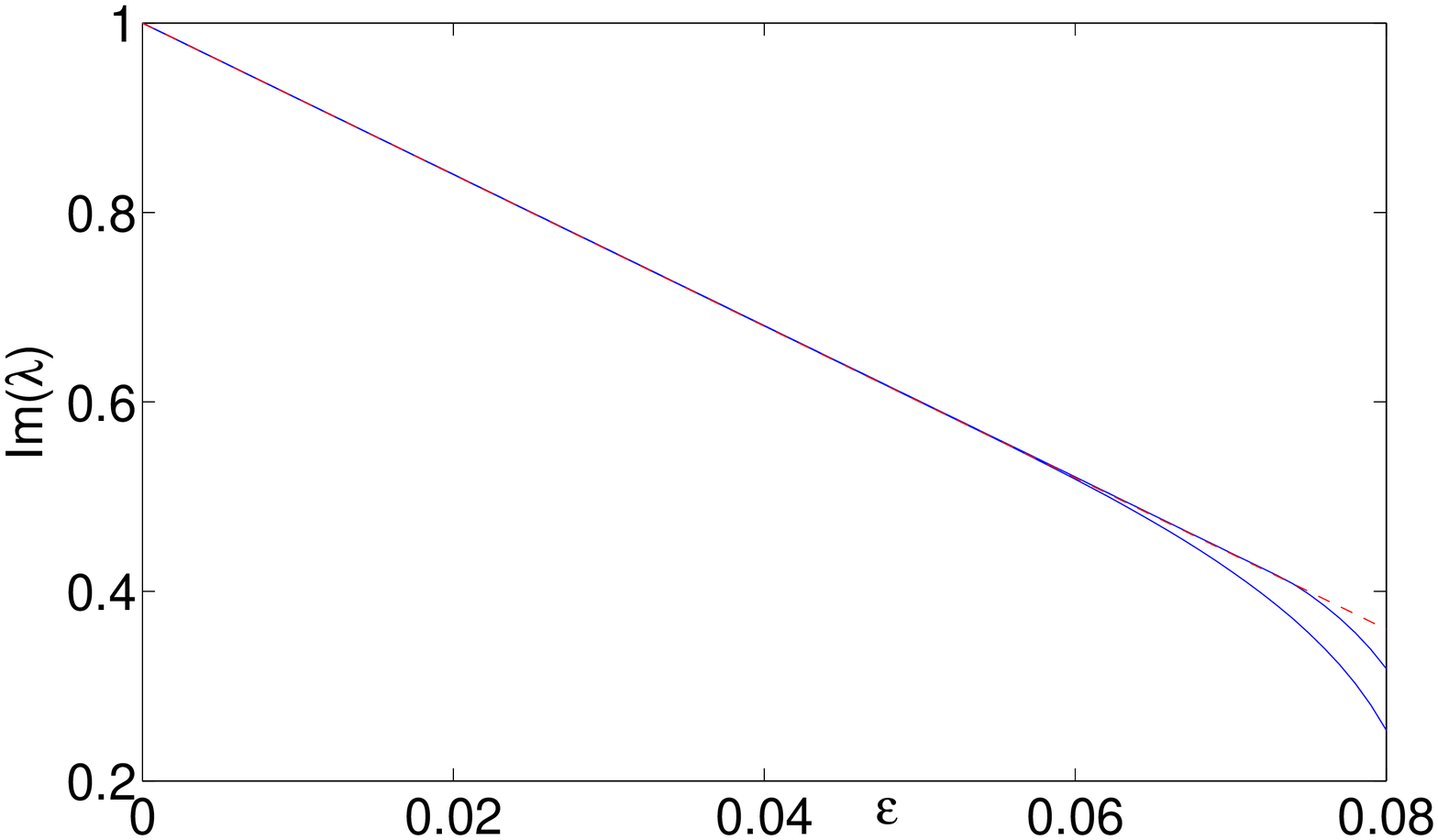} %
\includegraphics[width=0.45\textwidth]{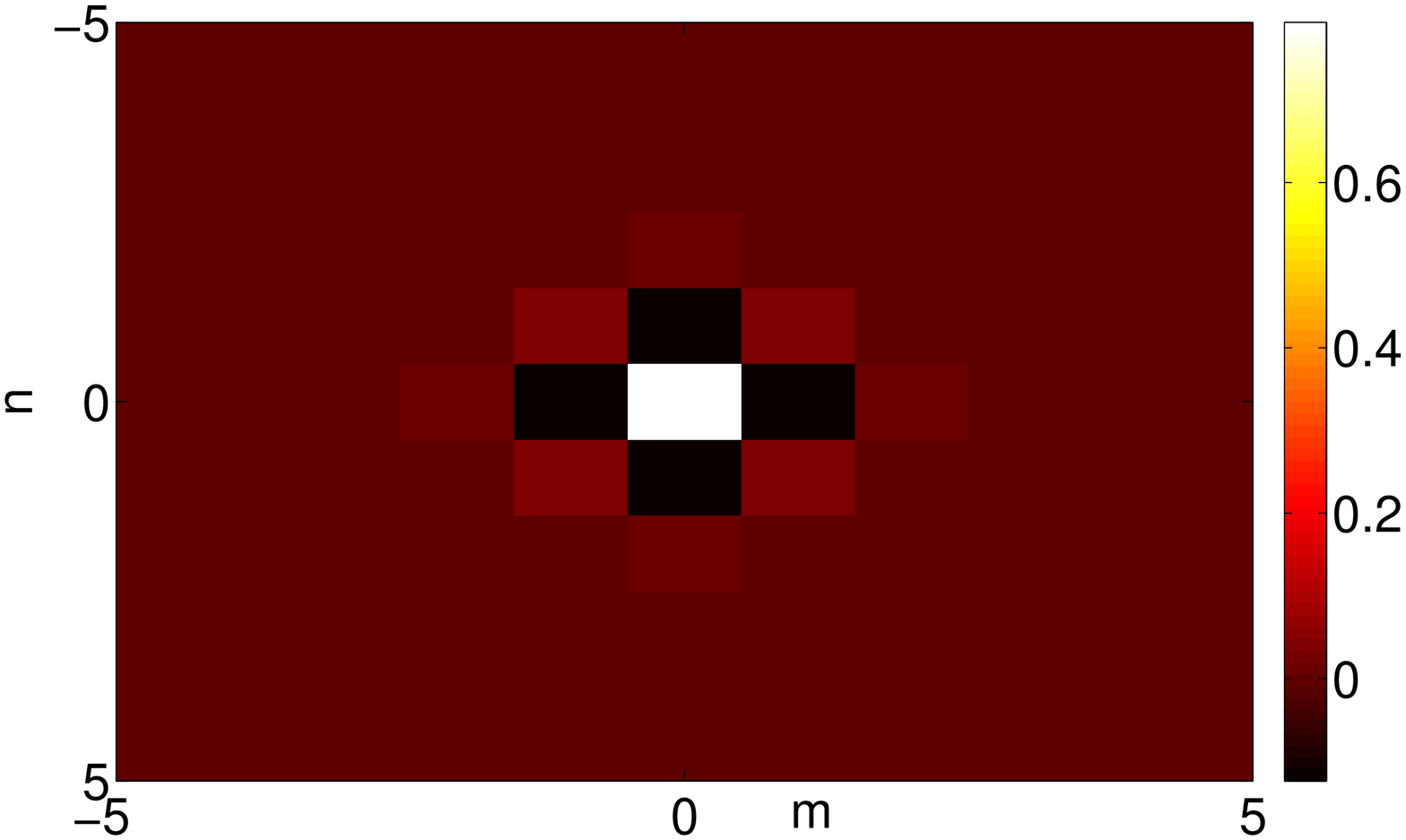}
\end{center}
\caption{(Color Online)
The branch corresponding to the single-site excitation. The left
panel shows eigenvalues bifurcating from the edge of the continuous spectrum
band (the first at $\protect\varepsilon =0.055$ and two more at a slightly
larger value of $\protect\varepsilon $), rapidly approaching the
spectral-plane's origin as $\protect\varepsilon \rightarrow 0.082$, the
value at which the present branch terminates.  Here and in similar
plots displayed below, the dependence of the edge of the continuous wave
band on $\protect\varepsilon $, $\protect\lambda =\pm (1-8\protect%
\varepsilon )i$, is shown by the red dashed line. 
The right panel shows the
profile of the branch at $\protect\varepsilon =0.08$.}
\label{new_fig1}
\end{figure}

In our numerical analysis, we first explore branches of stationary states
and their stability, and then proceed to simulations of the evolution of
perturbed solutions. The first localized state we consider in the AC limit
is the single-site one. This solution family is characterized by the
dependence of norm (\ref{N}) on the
coupling constant $\varepsilon $, which is
displayed in the left panel of Fig. \ref{new_fig1a} (the nearly linear shapes
of the dependences observed in this figure are explained by the small size
of the respective range of $\varepsilon $). Principal eigenvalues associated
with this branch, as well as a typical example of its profile (for $%
\varepsilon =0.08$), are shown in Fig.~\ref{new_fig1}. As indicated in the
previous section, throughout its existence region, this branch is stable,
with a single pair of eigenvalues at the origin. For this branch, multiple
pairs of eigenvalues bifurcate from the edge of the continuous-spectrum
band, $\lambda =\pm (1-8\varepsilon )i$: the first one bifurcates around $%
\varepsilon =0.055$, and the branch cannot be continued past $\varepsilon
=0.082$. It can be clearly seen from its profile close to this termination
point that it collides with a branch bearing a positive excitation at the
central site and a negative excitation at adjacent ones.

\begin{figure}[th]
\par
\begin{center}
\includegraphics[width=0.45\textwidth]{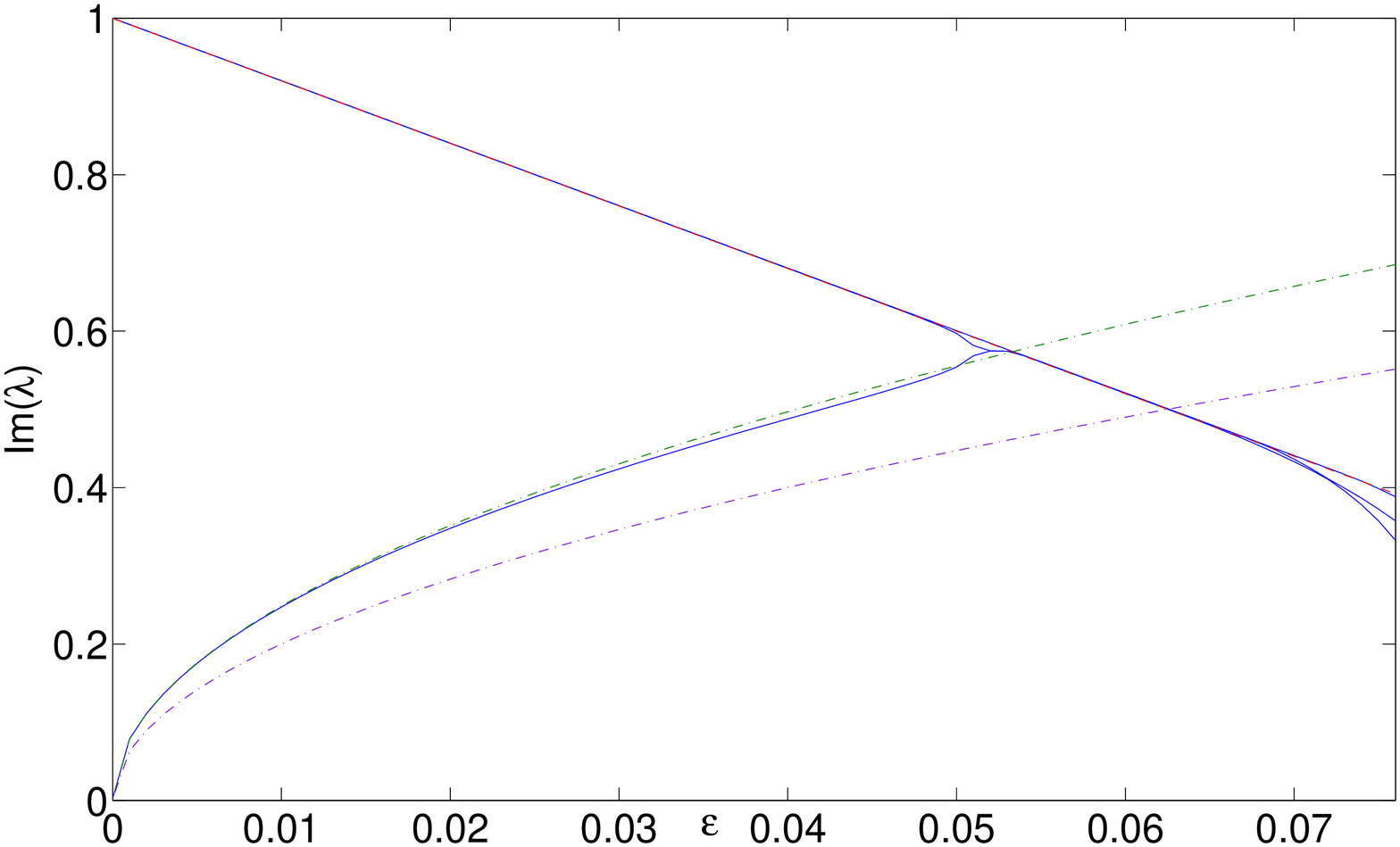} %
\includegraphics[width=0.45\textwidth]{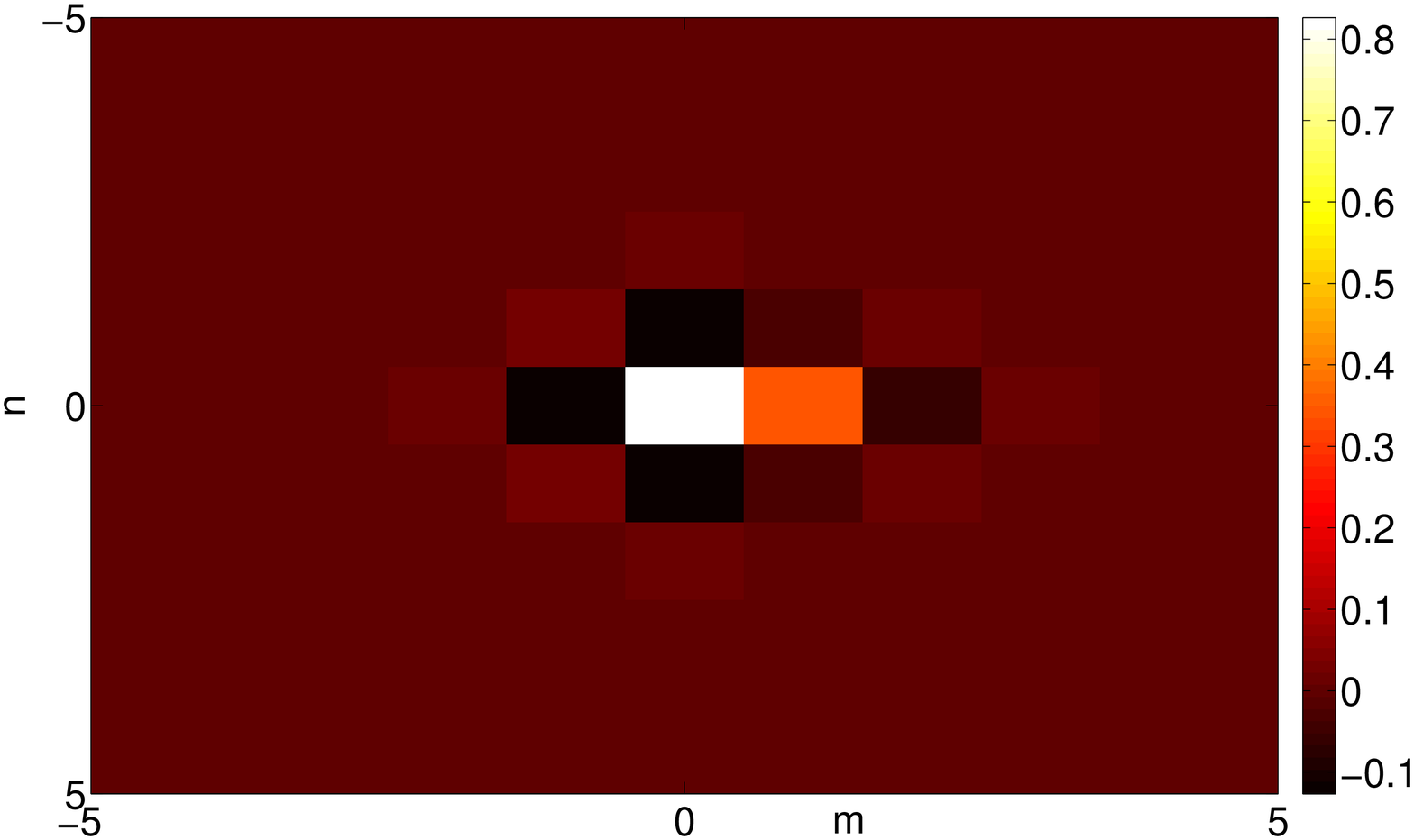}
\end{center}
\caption{(Color Online) In-phase configuration: 
the left panel shows the imaginary eigenvalue growing from the
origin, as per numerical results (blue solid line), according to the
analytical prediction (green dash-dotted line), and in the homogeneous model
(the lower dash-dotted line, obtained in an analytical form too).
Eigenvalues bifurcating from the edge of the continuous-spectrum band are
also shown by blue solid lines. The right panel displays an example of this
waveform for $\protect\varepsilon =0.079$.}
\label{new_fig2}
\end{figure}

\begin{figure}[!ht]
\par
\begin{center}
\includegraphics[width=0.45\textwidth]{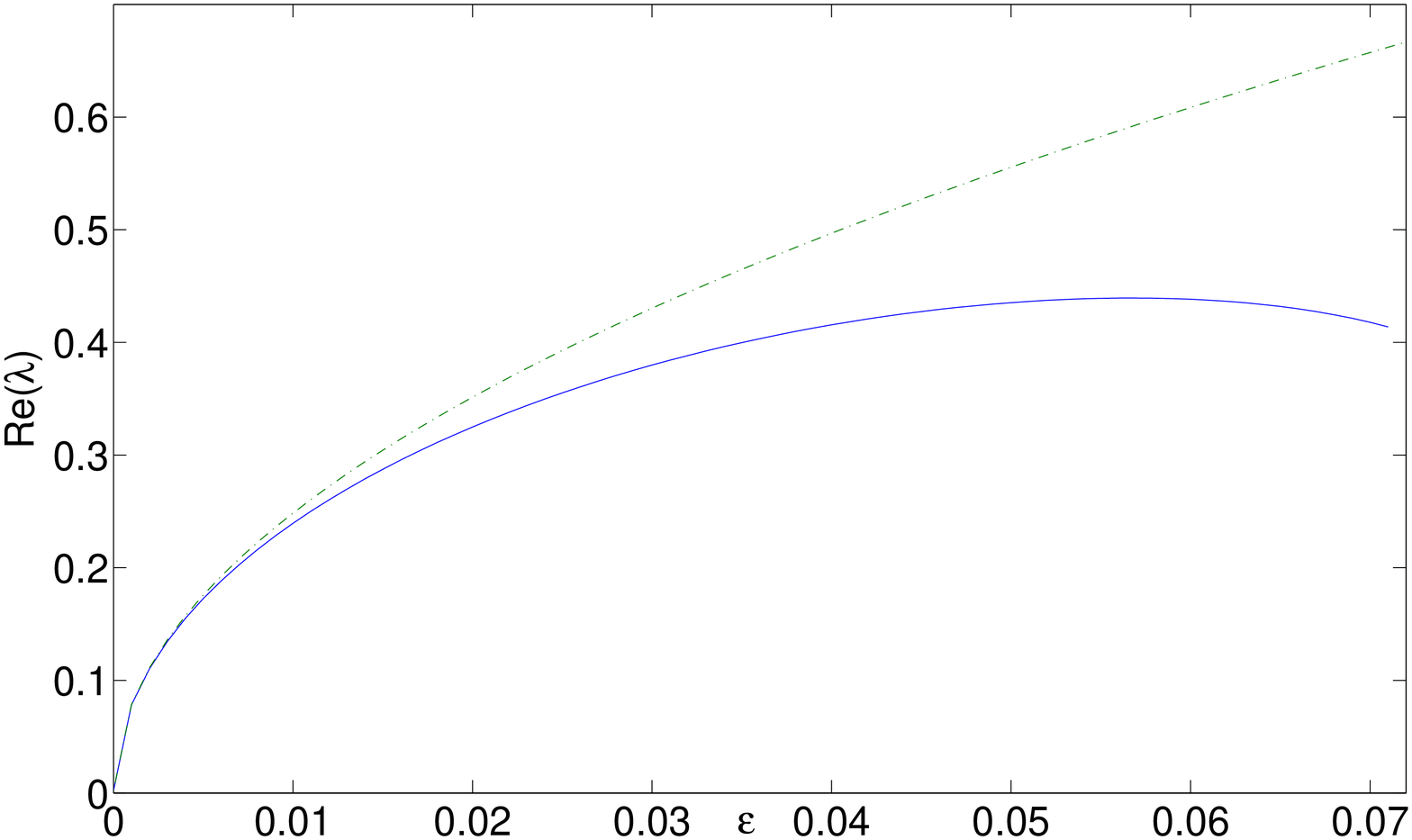} %
\includegraphics[width=0.45\textwidth]{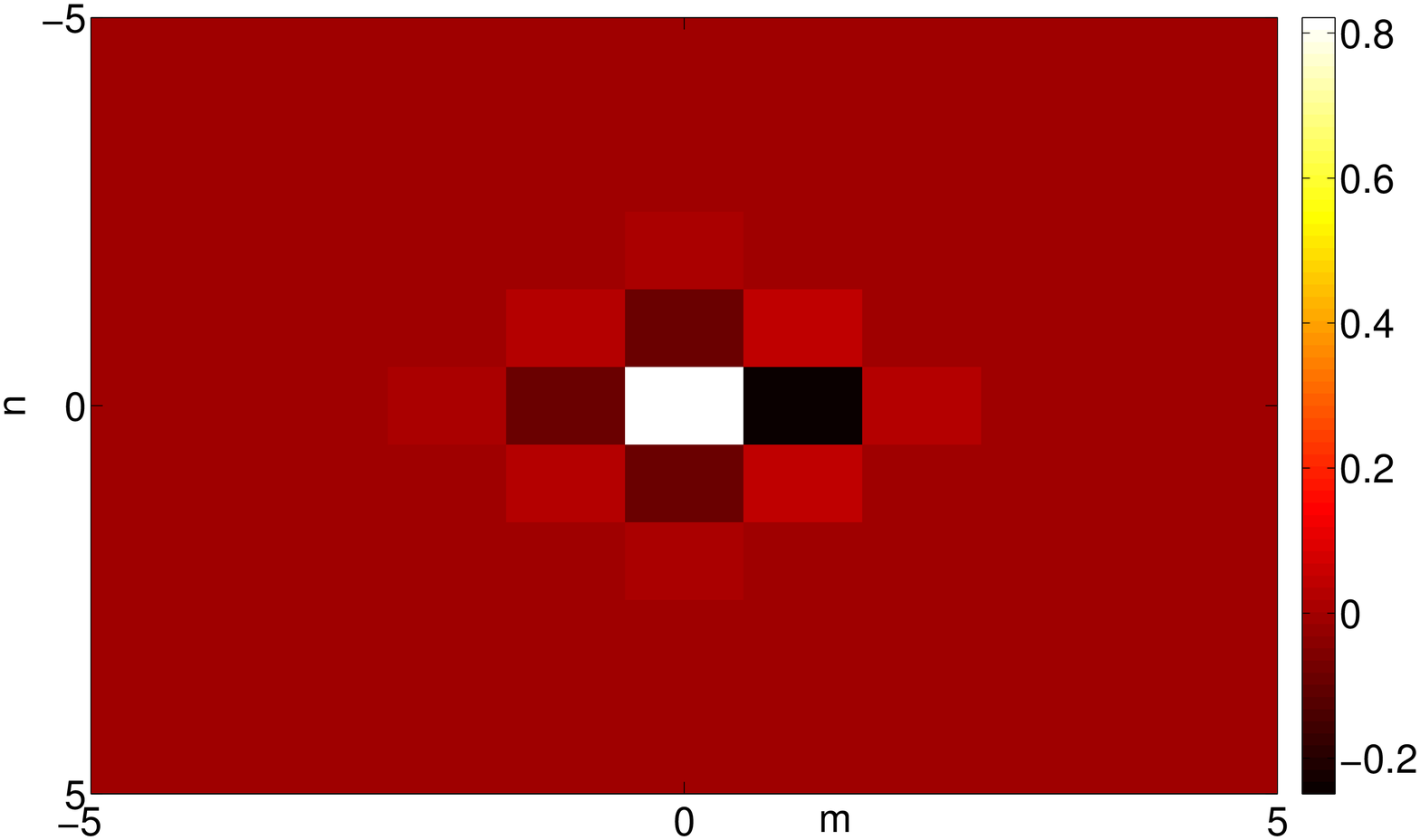}
\end{center}
\caption{(Color Online) 
Similar to the previous graph but for the out-of-phase
configuration. In this case, however, the eigenvalue pair bifurcating from
the origin moves to the real line and hence the real part of the relevant
eigenvalue is shown (blue solid line: numerical linear stability result,
green dash-dotted line: theory). The right panel shows the corresponding
waveform for $\protect\varepsilon=0.07$.}
\label{new_fig3}
\end{figure}

The next two branches we examine correspond to two-site excitations. The
in-phase and out-of-phase ones are shown by the middle
and right panels of Fig. \ref{new_fig1a}, and by Figs.~\ref{new_fig2} and~%
\ref{new_fig3}. In the former case, the eigenvalue bifurcating from the
origin is approximately $\pm 2.484\sqrt{\varepsilon }i$. It collides with
the band edge, $\pm (1-8\varepsilon )i$ at $\varepsilon =0.053$ or $0.052$,
according to the analytical approximation and numerical results,
respectively, which demonstrates a very good agreement between the
two in the
prediction of the threshold for the oscillatory instability arising for this
branch, as well as for the entire $\varepsilon $-dependence of the
eigenvalue pair. In Fig.~\ref{new_fig2} we also show, by means of the lower
(magenta) curve, $\mathrm{Im}(\lambda )=\pm 2\sqrt{\varepsilon }i$, the
analytical prediction for the homogeneous system, with $g(m,n)=1$. 
We note that in the inhomogeneous model the eigenfrequency pair
grows more rapidly, thus leading to an instability at a lower value of the
coupling, than in its homogeneous counterpart. The branch is unstable past
the point of $\varepsilon =0.052$, and for $\varepsilon >0.065$ further
eigenvalue pairs bifurcate off of the continuous spectrum, their collision with
the origin leading to the termination of the branch at larger values of $%
\varepsilon $; in the right panel of the figure, the branch is shown for $%
\varepsilon =0.079$.

In the case of the out-of-phase two-site excitation, as seen in Fig.~\ref%
{new_fig3}, the linearization around the solution produces a real eigenvalue
pair predicted to be $\lambda =\pm 2.484\sqrt{\varepsilon }$, which is
reasonably accurate for small $\varepsilon $. For larger values of $%
\varepsilon $, higher-order terms apparently take over, pulling the
eigenvalue back to the origin (nevertheless, the instability is present at
all the values of $\varepsilon $ that we considered). A typical example of
the profile of the discrete mode is shown in the right panel of Fig.~\ref%
{new_fig3} for $\varepsilon =0.07$. The profile suggests that the solution
collides with the single-site one, and with the above-mentioned cross-shaped
solution with four negatively excited sites around the central one.
Therefore, the present mode represents one of the four asymmetric branches
--the other three arise by rotating the present one by $\pi /2$, $\pi $ and $%
3\pi /2$ (see the right panel of Fig.~\ref{new_fig3})--, 
which are generated by a pitchfork bifurcation.

\begin{figure}[th]
\par
\begin{center}
\includegraphics[width=0.45\textwidth]{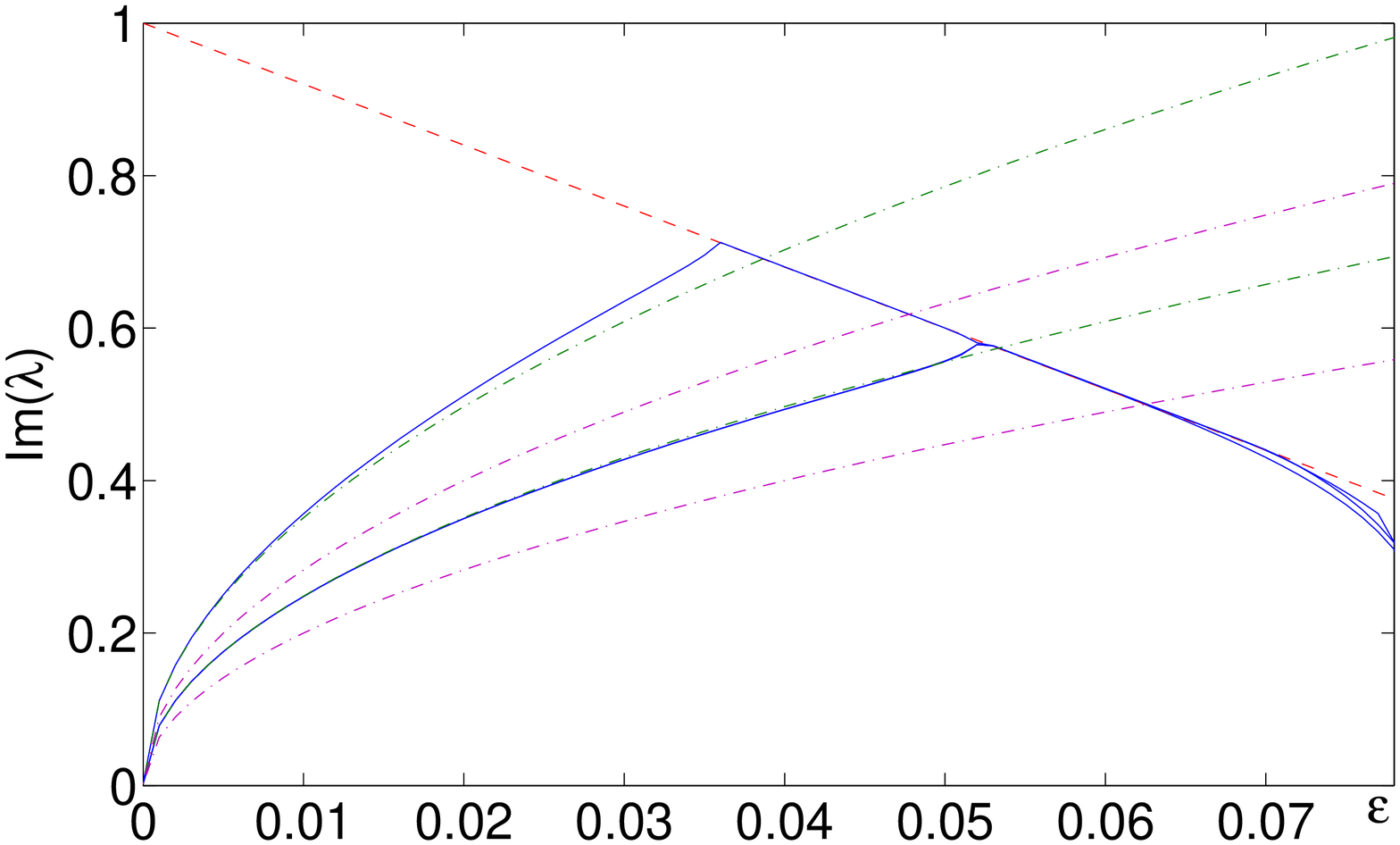} %
\includegraphics[width=0.45\textwidth]{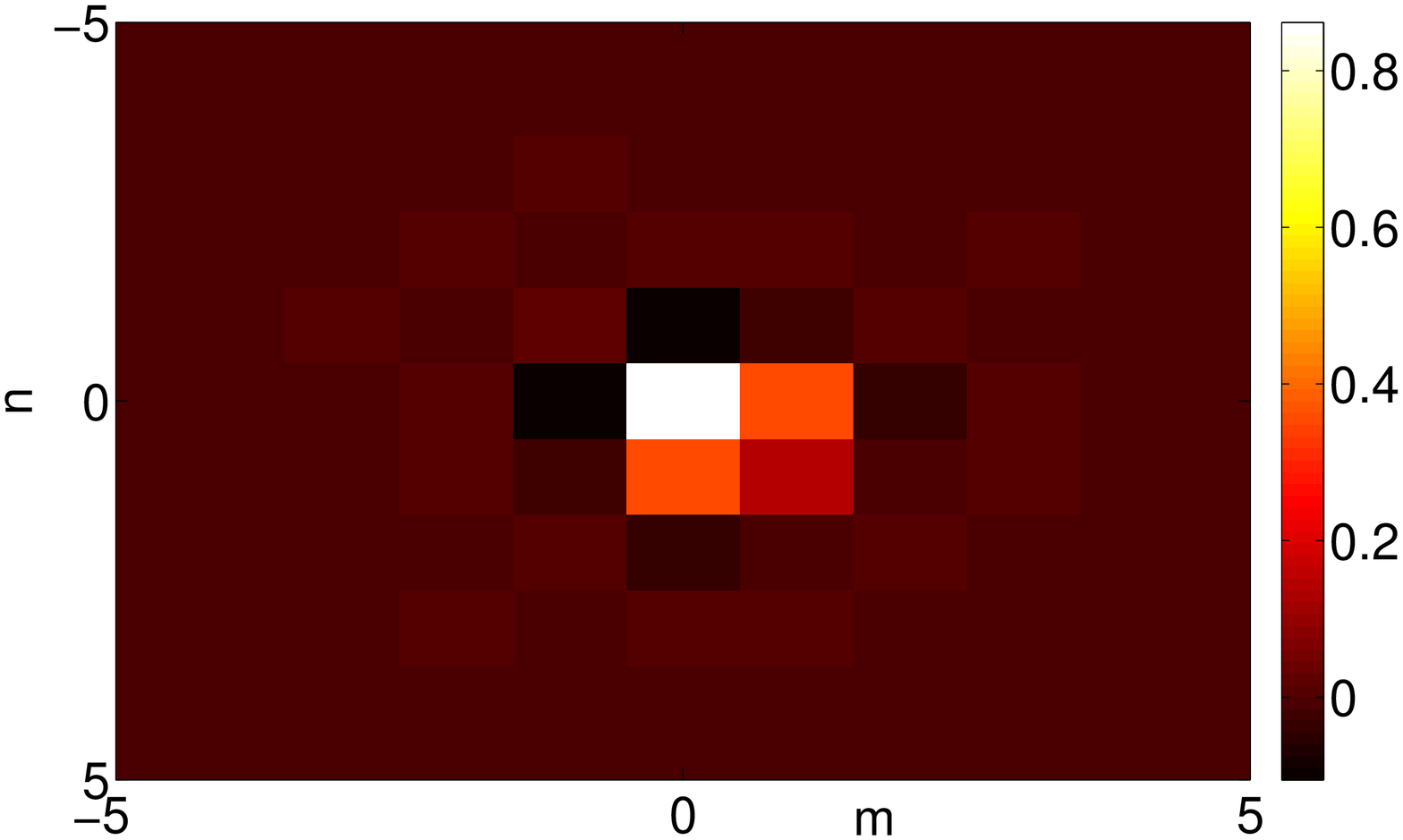}
\end{center}
\caption{(Color Online) 
Four-site, in-phase excitation: in the left panel of the figure
three pairs of imaginary eigenvalues bifurcating from the origin are shown
by the blue solid line (the lower parabolic line corresponds to a double
pair). The corresponding analytical prediction is shown by the green
dash-dotted line, while the prediction for the homogeneous model is
presented by the magenta dash-dotted line. The edge of the
continuous-spectrum band is shown by the red dashed line. The right panel
shows the configuration for $\protect\varepsilon =0.075$.}
\label{new_fig4}
\end{figure}

\begin{figure}[th]
\par
\begin{center}
\includegraphics[width=0.45\textwidth]{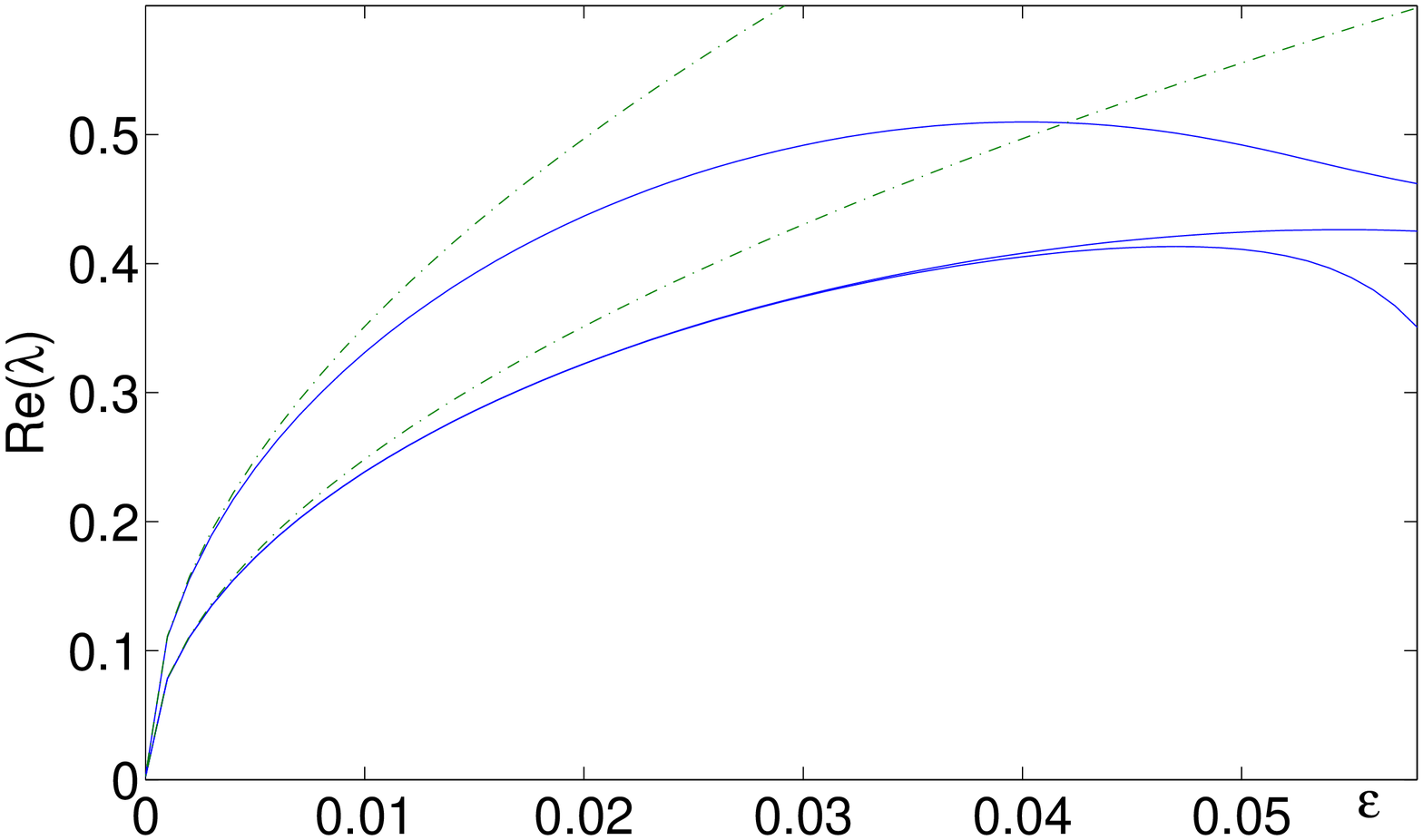} 
\includegraphics[width=0.45\textwidth]{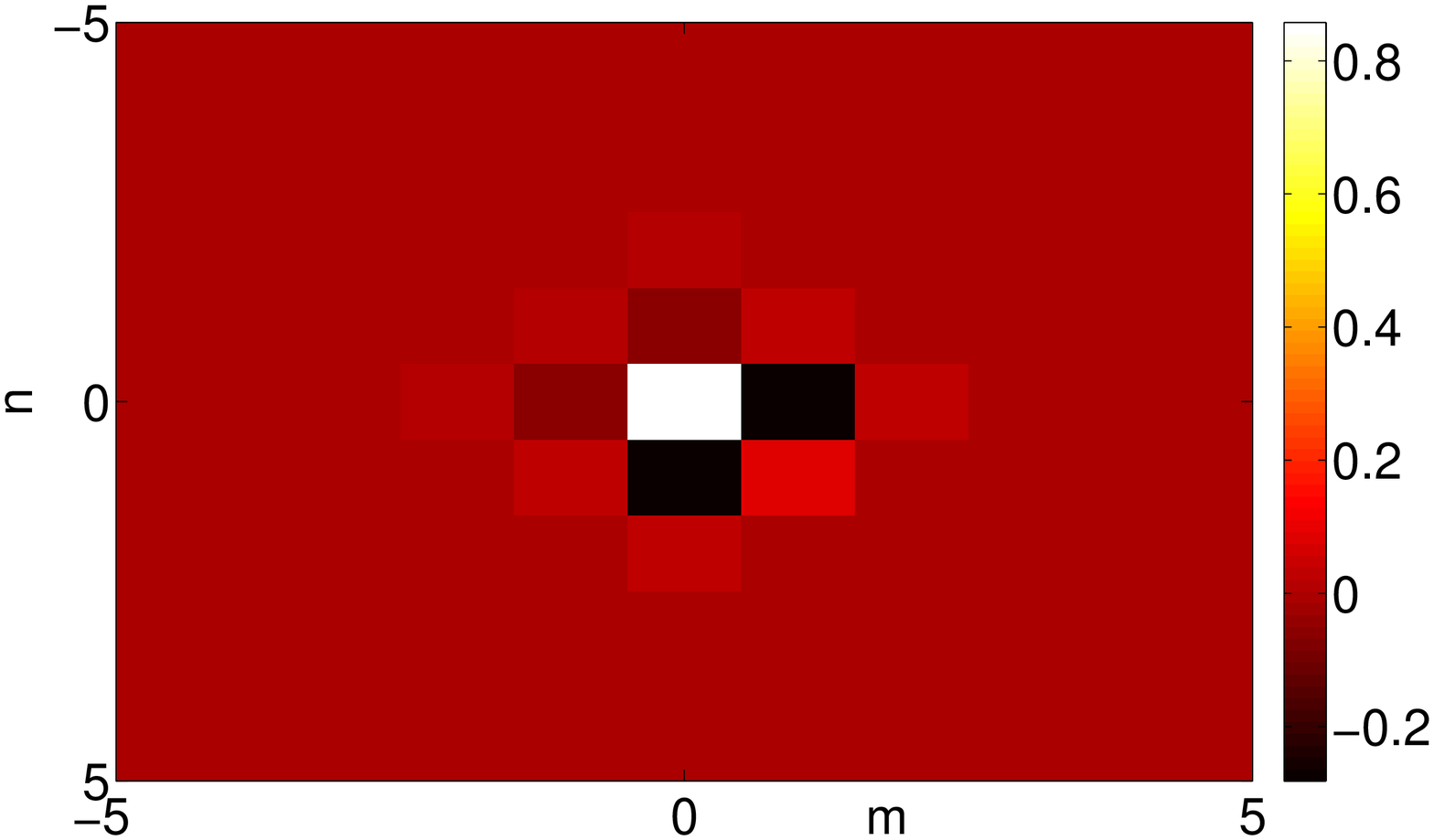}
\end{center}
\caption{(Color Online)
Similar to the previous figures, but now for the four-site,
out-of-phase excitation. The left panel shows the numerical finding (the
blue solid) and the analytical prediction (the green dash-dotted line) for
the three pairs bifurcating from the origin towards the real axis, rendering
the configuration highly unstable. A typical example of the configuration
profile for $\protect\varepsilon =0.055$ is shown in the right panel.}
\label{new_fig5}
\end{figure}

Examining now the four-site excitations in the framework of the analysis
based on Eq.~(\ref{d2dnls12}), we conclude that asymmetric configurations
always bear a number of instabilities. 
[By ``asymmetric" here,
we mean configurations other than the in-phase one, in which all phases
of the four excited sites are the same e.g. $(0,0,0,0)$, and the
out-of-phase one, in which the phases alternate between $0$ and $\pi$, e.g. $%
(0,\pi,0,\pi)$ for the four sites of the square. Any other phase
combination, e.g. $(0,\pi,\pi,0)$, $(0,0,\pi,0)$ etc. is considered
asymmetric.]
For demonstration purposes, we
restrict our considerations here 
to the two most symmetric examples, namely the
in-phase state shown in Fig.~\ref{new_fig4}, and the out-of-phase one in
Fig.~\ref{new_fig5}. In the case of the four-site in-phase excitation, there
are three eigenvalue pairs bifurcating from the spectral-plane's origin,
whose behavior is determined by Eq.~(\ref{d2dnls12}). One of the three pairs
is predicted to have eigenvalues $\lambda =\pm 3.513\sqrt{\varepsilon }i$,
and two others to have $\lambda =\pm 2.484\sqrt{\varepsilon }i$. As a
result, by setting these eigenvalues equal to the edge of the
continuous-spectrum band, $(1-8\varepsilon )i$, we can predict the onset of
instabilities at $\varepsilon =0.0386$ (the first one), and at $\varepsilon
=0.053$ (a pair of additional ones). Numerically these instabilities are
found to occur, respectively, at $\varepsilon =0.036$ and at $\varepsilon
=0.053$, in very good agreement with the 
theoretical predictions. Generally, in the
present case of the four-site modes, we again observe good agreement between
the analytical predictions for the eigenvalues and their numerical
counterparts. In addition, we point out that here too the eigenvalues move
more rapidly along the imaginary axis than their counterparts in the
homogeneous model (shown by magenta dash-dotted lines in Fig.~\ref{new_fig4}%
). The configuration is generically unstable for $\varepsilon >0.036$, and a
typical example of this mode is shown for $\varepsilon =0.075$ in the right
panel of the figure.

In the case of the four-site, out-of-phase (between adjacent sites)
excitation, the analysis produces three real eigenvalue pairs, bifurcating
from the origin as $\varepsilon $ increases. The dominant one is predicted
to be $\lambda =\pm 3.513\sqrt{\varepsilon }$, while two more correspond to $%
\lambda =\pm 2.484\sqrt{\varepsilon }$. Here again, as is typical for
configurations with real eigenvalue pairs, the agreement between the
analytical prediction and numerical results is good for small coupling
strengths but progressively deteriorates as the coupling grows stronger and
the configuration develops towards its collision/bifurcation with other
states. The state is found to be unstable for all values of the coupling. A
typical example of the state for $\varepsilon =0.055$ is shown in the right
panel of Fig.~\ref{new_fig5}.

\begin{figure}[th]
\par
\begin{center}
\includegraphics[width=0.45\textwidth]{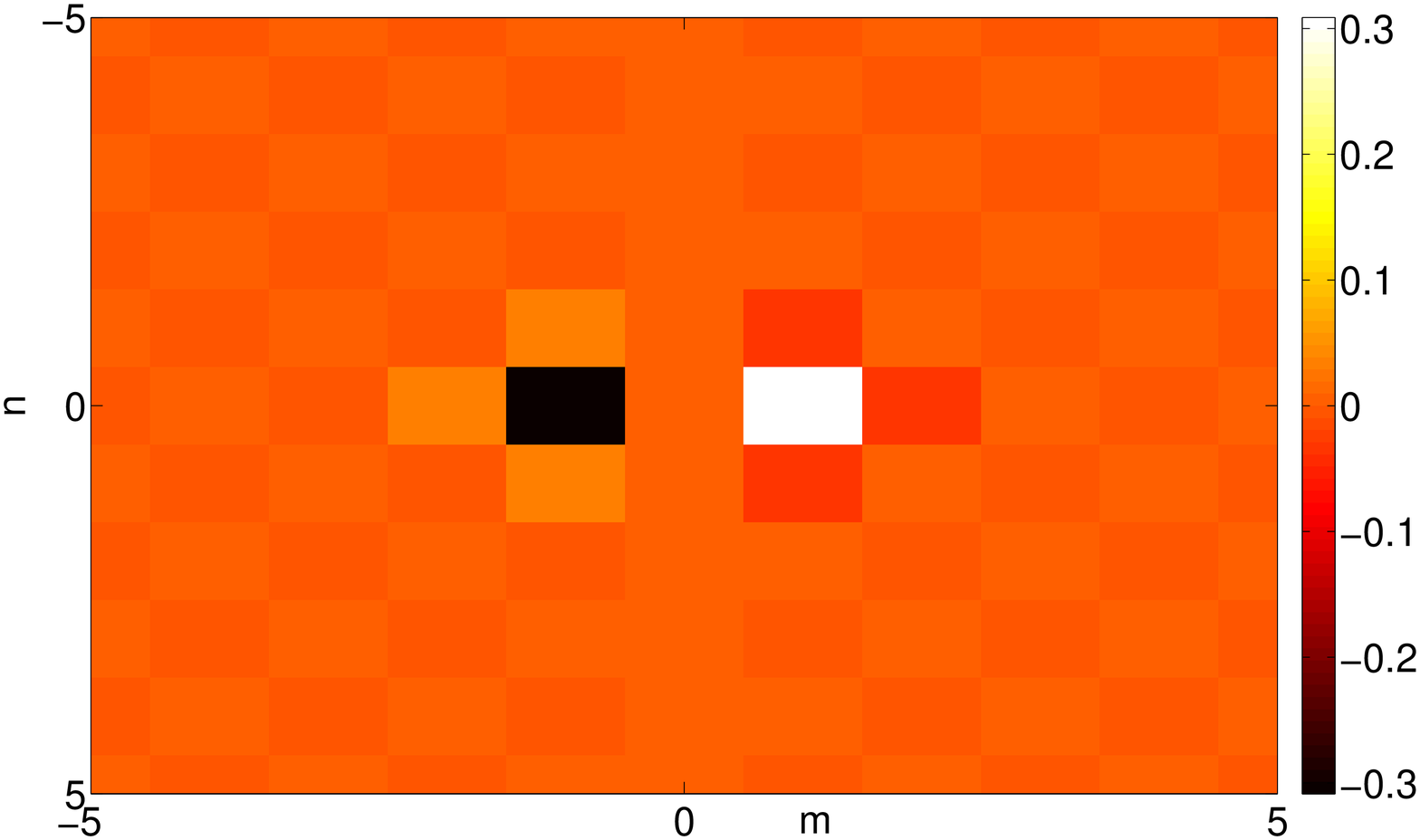} 
\includegraphics[width=0.45\textwidth]{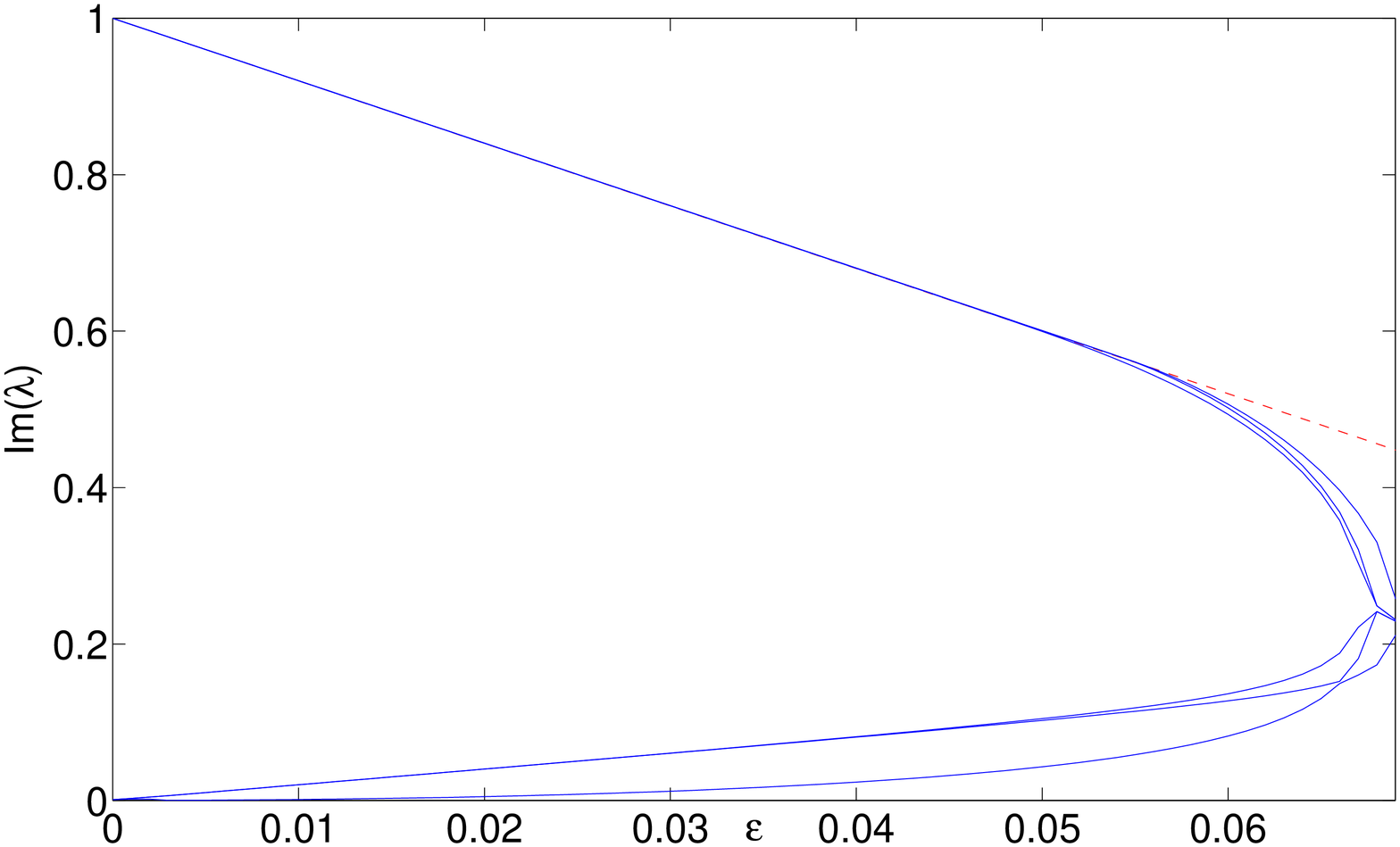}
\par
\includegraphics[width=0.45\textwidth]{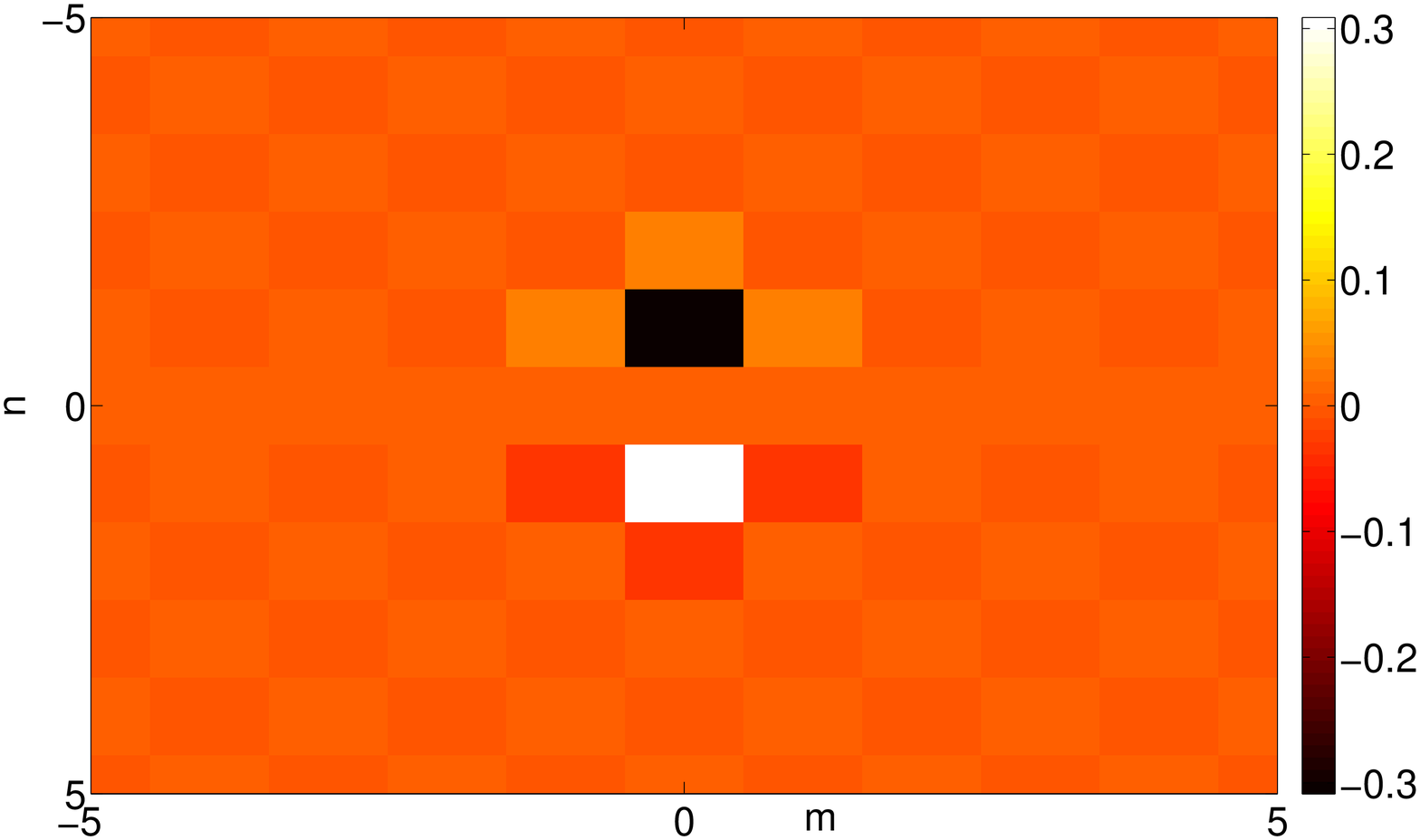} 
\includegraphics[width=0.45\textwidth]{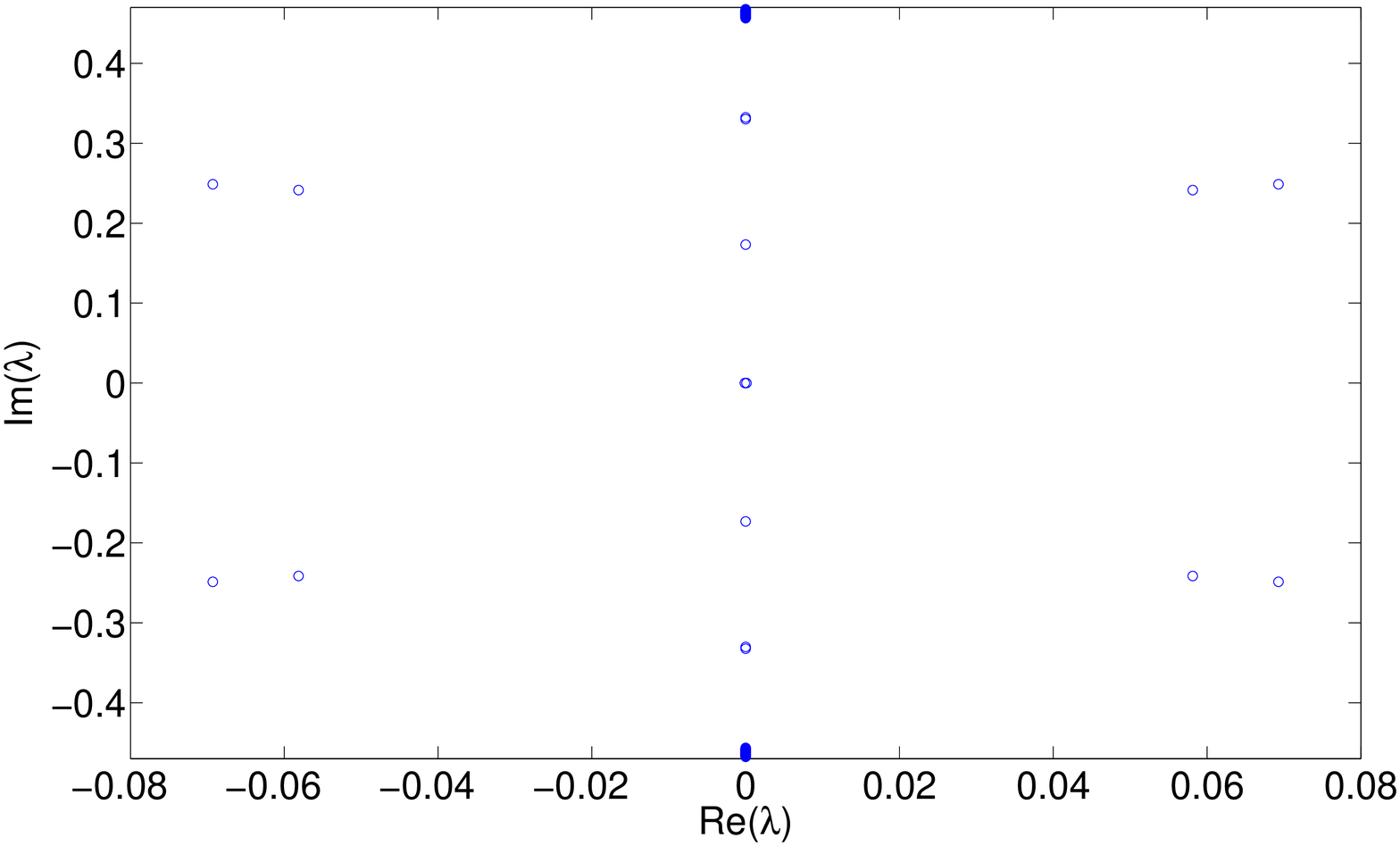}
\end{center}
\caption{(Color Online) 
The left panels of the figure illustrate the real (top) and
imaginary (bottom) parts of the spatial distribution of a two-dimensional
vortex cross. The phases of the four excited sites are $0$, $\protect\pi /2$%
, $\protect\pi $ and $3\protect\pi /2$, so that a phase circulation of $2%
\protect\pi $ is achieved when moving along a contour surrounding the mode's
pivot. The solution is shown for $\protect\varepsilon =0.068$, and its
corresponding spectral plane is displayed in the bottom right panel. The top
right panel illustrates the $\mathcal{O}(\protect\varepsilon )$ (or weaker;
see the smallest eigenvalue pair) dependence for small $\protect\varepsilon $
of the eigenvalue pairs bifurcating from the origin. It is the collision of
these pairs with the ones bifurcating from the band edge (which is depicted,
as before, by the dashed red line), that leads to the instability at $%
\protect\varepsilon \geq 0.068$.}
\label{new_fig6}
\end{figure}

We now turn to the examination of the only genuinely complex state
considered here, namely the discrete ``vortex cross"; see Fig.~\ref{new_fig6}%
. While we were able to identify and continue this type of state in the
symmetric pattern illustrated in the figure, it is worth noting that when
we attempted to construct a similar configuration based on the square of the
four-site excitations shown in Figs.~\ref{new_fig4} and \ref{new_fig5}, we
were unable to continue it to finite couplings. It is unclear whether this
type of vortex-square mode exists; this is
 a subject which merits further
investigation.

In the case of the vortex crosses of Fig.~\ref{new_fig6}, however, there
are three eigenvalue pairs that bifurcate from the origin along the
imaginary axis, attesting to the stability of the structure for small $%
\varepsilon $. Importantly, these eigenvalues scale $\propto \varepsilon $,
rather than $\sqrt{\varepsilon }$, and can only be captured at the second
order of perturbation theory, which is not considered here. As these
eigenvalue pairs move along the imaginary axis, further pairs bifurcate from
the edge of the continuous-spectrum band at $\lambda =\pm (1-8\varepsilon )i$%
, starting at approximately $\varepsilon =0.05$. As $\varepsilon $
increases, these pairs approach each other and eventually collide around $%
\varepsilon =0.068$ (shown in the right panels of Fig.~\ref{new_fig6}),
rendering the branch of solutions unstable past this critical point. This
happens because collisions of the former eigenvalue pair, bifurcating from $%
0 $, with the latter one, which bifurcates from the band edge, give rise
to complex quartets and oscillatory instabilities.

Finally, we consider the extended state in which all the sites of the
lattice are excited in accordance to the TFA, $v_{m,n}=\sqrt{\mu /g(m,n)}$,
cf. Eq. (\ref{TF}). First, assuming that the coupling constant 
$\varepsilon $ is small, Eqs.~(\ref{TF}) and (\ref{g}) readily
yield the leading order correction to the TFA, which, by itself, corresponds to 
$\varepsilon =0$ (recall the chemical potential is fixed as $\mu =1$):
\begin{equation}
v_{m,n}\approx v_{m,n}^{(0)}+\varepsilon v_{m,n}^{(1)}=e^{-\left(
|m|+|n|\right) }-\left( \varepsilon /2\right) \Delta _{2}\left( e^{-\left(
|m|+|n|\right) }\right) .  \label{epsilon}
\end{equation}
%
In particular, the accordingly predicted amplitude of the extended
mode, at $m=n=0$, is
\begin{equation}
A_{\max }=1-2\left( 1-e^{-1}\right) \varepsilon \approx 1-1.264\varepsilon.
\label{Amax}
\end{equation}
%

\begin{figure}[th]
\par
\begin{center}
\includegraphics[width=0.45\textwidth]{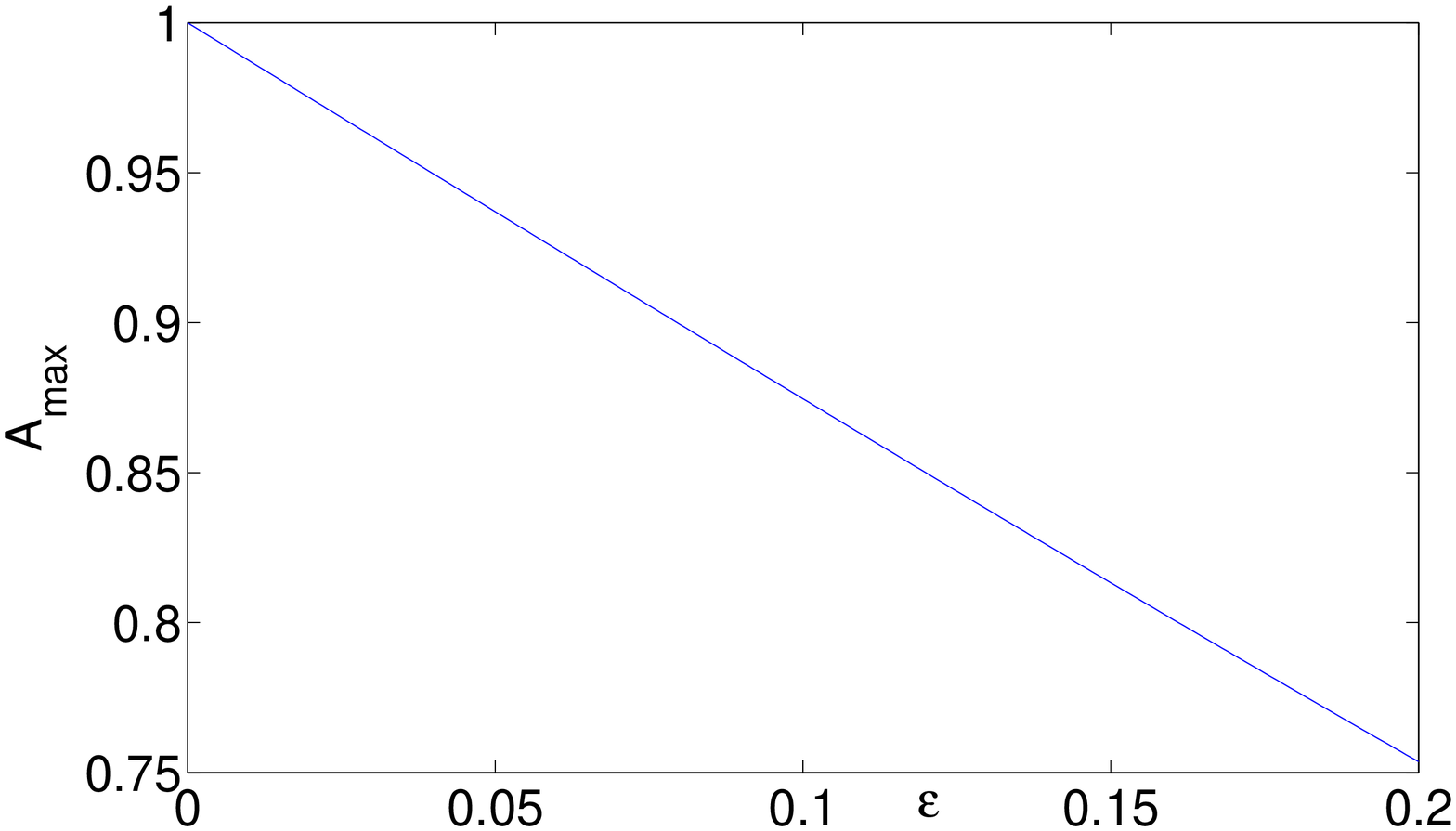} %
\includegraphics[width=0.45\textwidth]{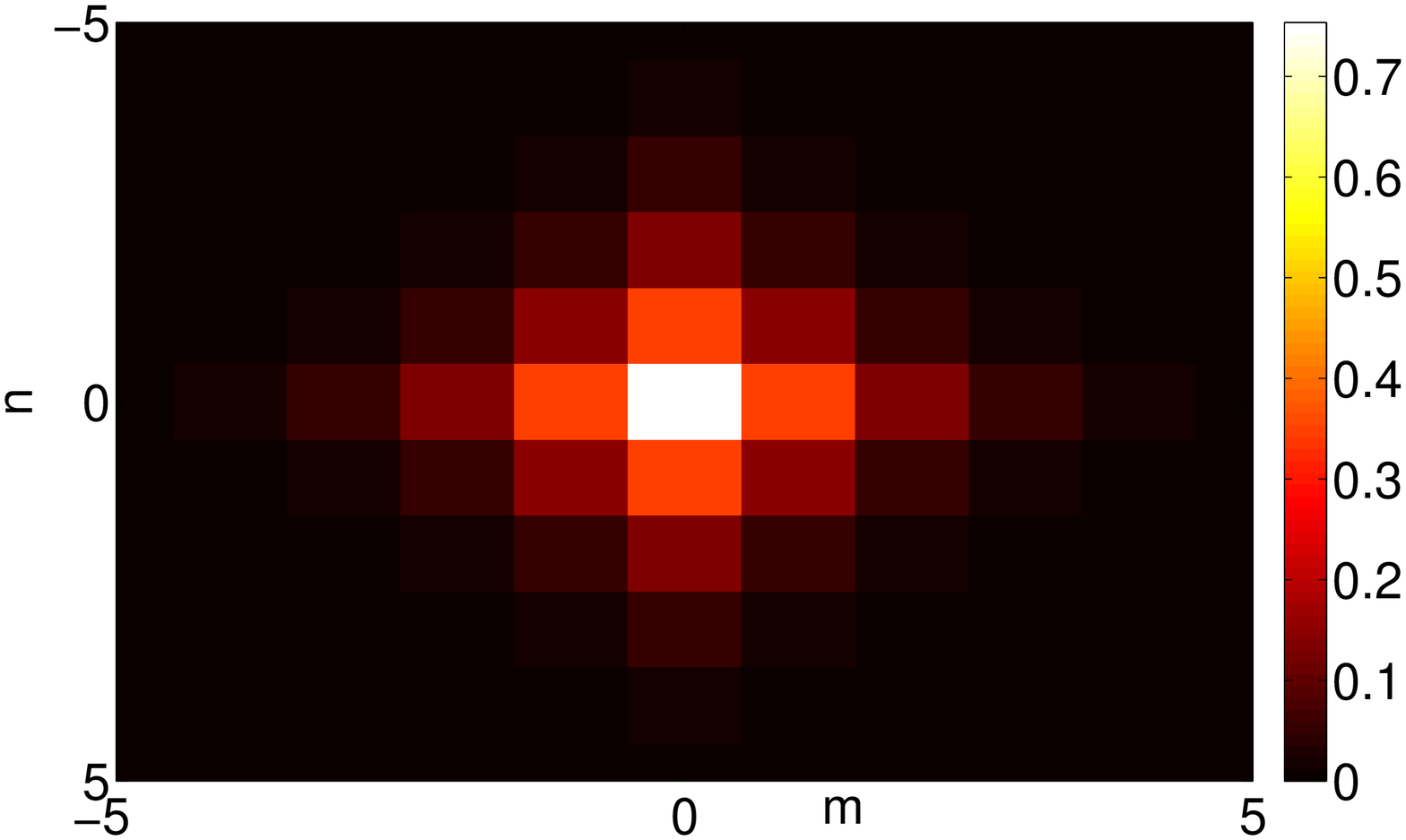}
\end{center}
\caption{(Color Online)
The left panel of the figure shows the dependence of the amplitude
of the extended solution as a function of $\protect\varepsilon $, while the
right panel displays a typical profile of the extended mode for $\protect%
\varepsilon =0.2$.}
\label{new_fig7}
\end{figure}

By means of our numerical continuation, it was possible to follow this
solution for all the values of the coupling that we considered, up to 
$\varepsilon =0.2$. As  can be seen in the left panel of the figure,
the dependence of the amplitude of the solution on the coupling constant is
almost exactly approximated by $A_{\max }=1-(5/4)\varepsilon $,
i.e., the perturbative result (\ref{Amax}) predicts the dependence very
accurately. An example of a numerically found profile of the mode is shown
in the right panel of Fig.~\ref{new_fig7} for $\varepsilon =0.2$.
Furthermore, the numerical analysis has demonstrated that the solution is
stable throughout its entire existence interval (up to $\varepsilon =0.2$;
the numerical solution was not extended to large values of $\varepsilon $).
Notice that here the constraint due to potential collision of eigenvalues
stemming from the origin and from the continuous spectrum does \textit{not%
} exist, as actually all eigenvalue pairs bifurcate from the origin along
the imaginary axis.

\subsection{Evolution of unstable modes}

\begin{figure}[th]
\par
\begin{center}
\includegraphics[width=0.45\textwidth]{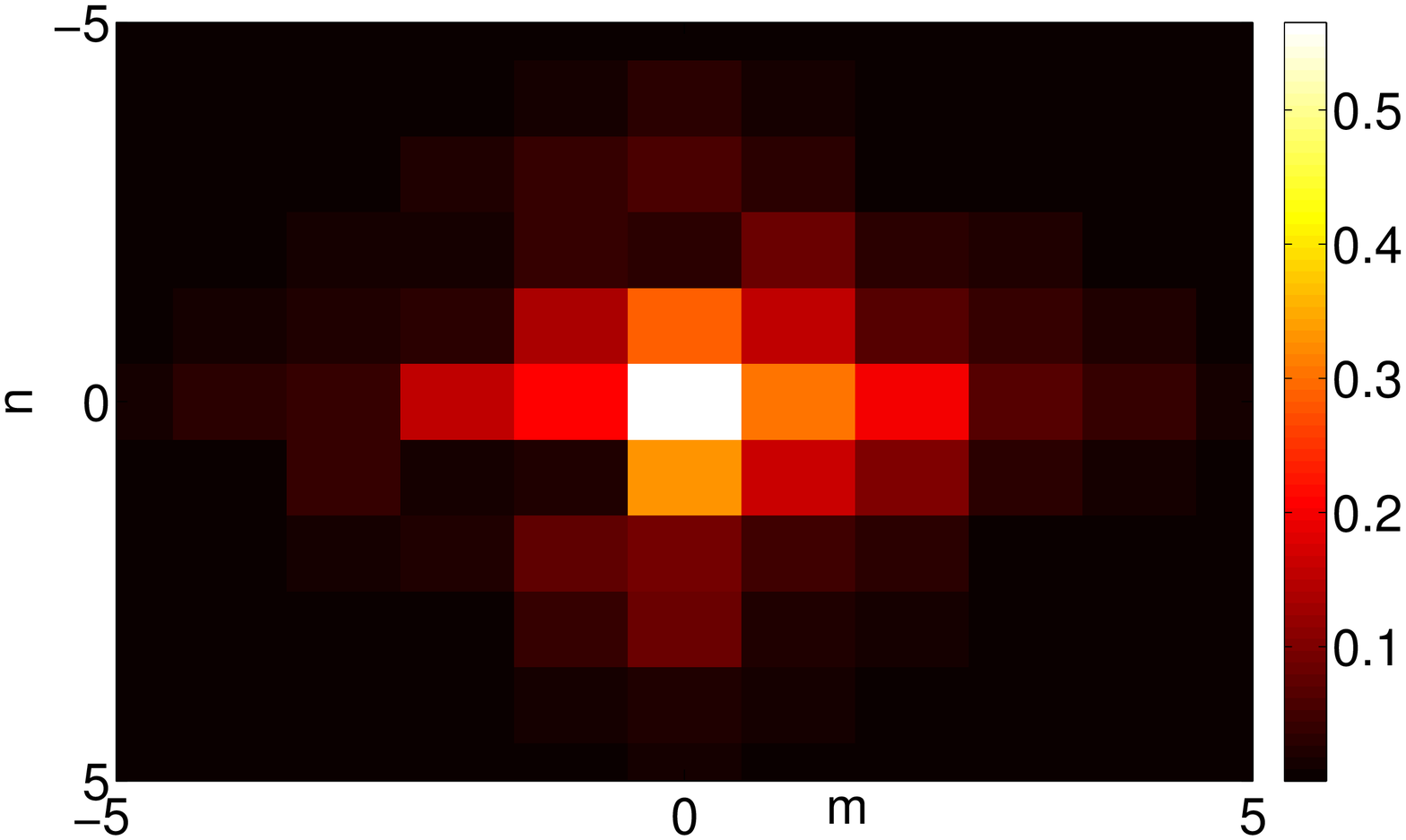} %
\includegraphics[width=0.45\textwidth]{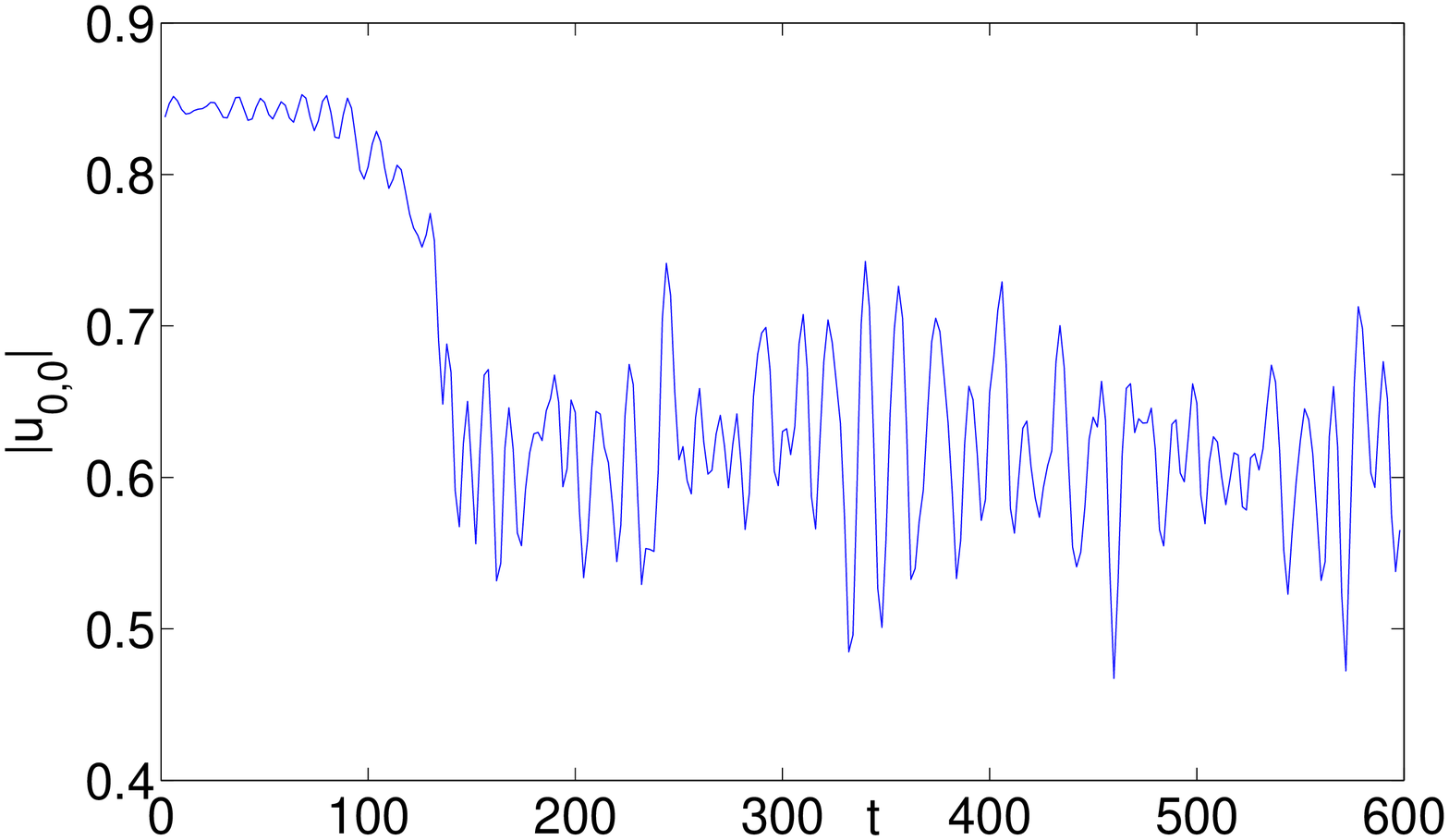}
\par
\includegraphics[width=0.45\textwidth]{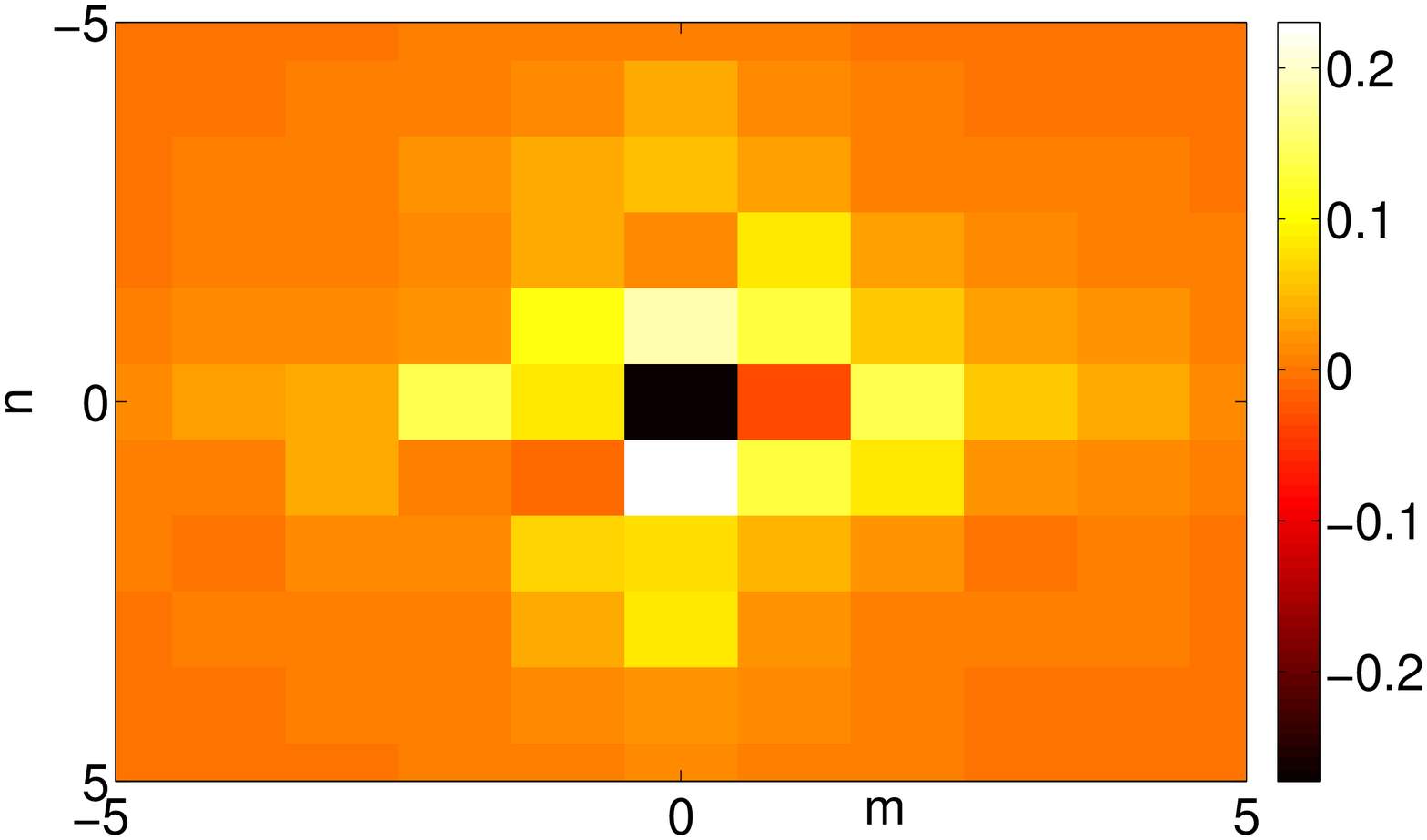} %
\includegraphics[width=0.45\textwidth]{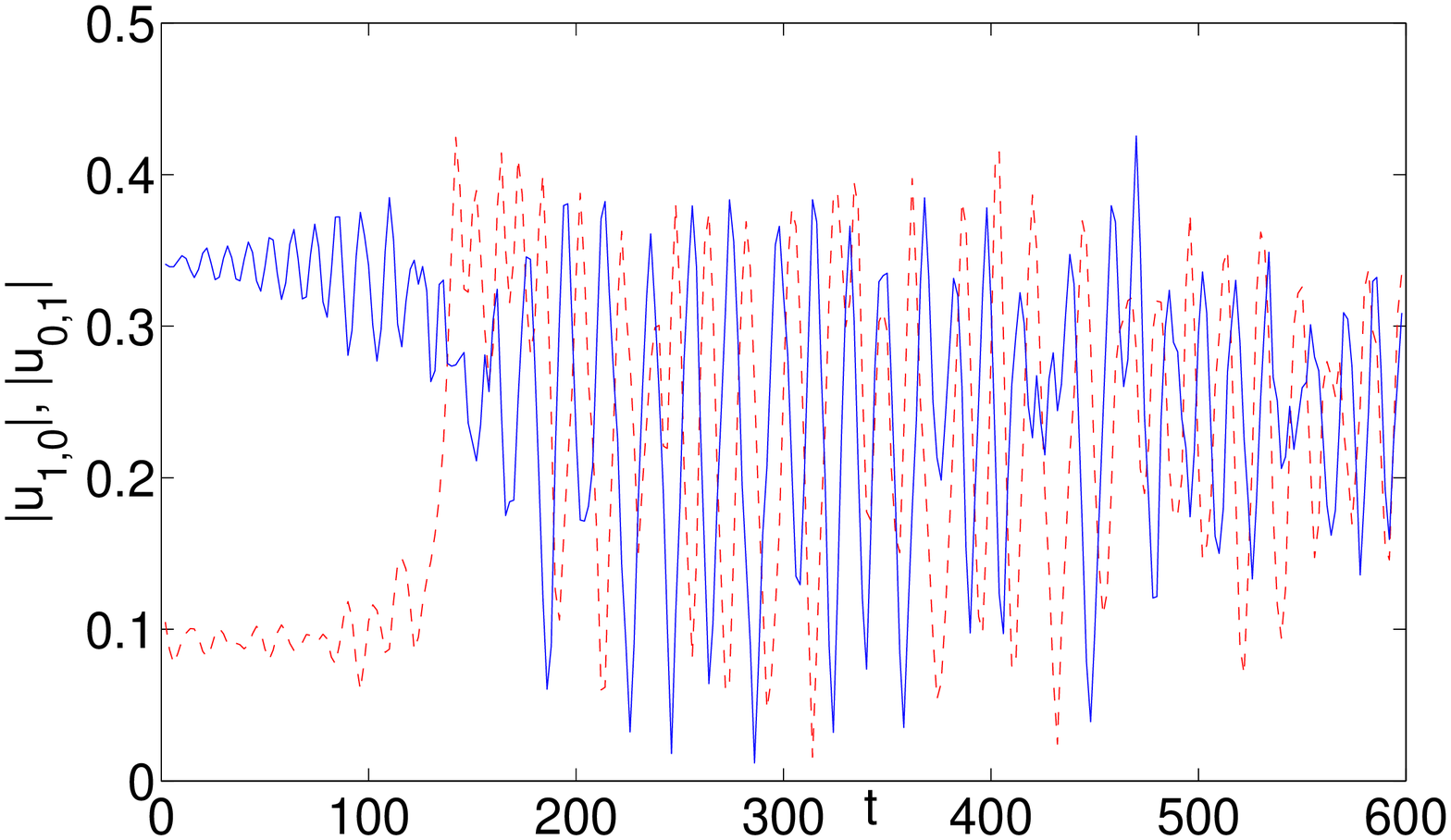}
\end{center}
\caption{(Color Online)
Evolution of the unstable two-site, in-phase mode at $\protect%
\varepsilon =0.079$. The top left panel shows the final profile of the
absolute value of the discrete wave field in the final state at $t=600$,
while the bottom left panel shows the difference between absolute values of
the top left profile and the initial one. The right panels show the absolute
value at the central site (top) and at two adjacent ones (bottom); the blue
solid line corresponds to the initially excited $(1,0)$ site, and the red
dashed line to the $(0,1)$ site. }
\label{new_fig7b}
\end{figure}

We now turn to direct numerical simulations of different unstable states.
Given that 
the single-site excitation is stable throughout its domain of existence, we
start with the two-site in-phase configuration in Fig.~\ref{new_fig7b}.
The top left panel of the figure shows the final profile of the solution
produced by simulations at $t=600$, for $\varepsilon =0.079$. The initial
condition is the mode from Fig.~\ref{new_fig2}, weakly perturbed by a
multiplicative small-amplitude random perturbation, intended to initiate the
instability. The bottom left panel shows the difference between the initial
and final profiles, illustrating how the instability expands across the
solution. The right panels of the figure show the evolution at the central and
adjacent sites, corroborating the same picture. Also evident in the latter
is the oscillatory character of the instability associated with this
solution.

\begin{figure}[th]
\par
\begin{center}
\includegraphics[width=0.45\textwidth]{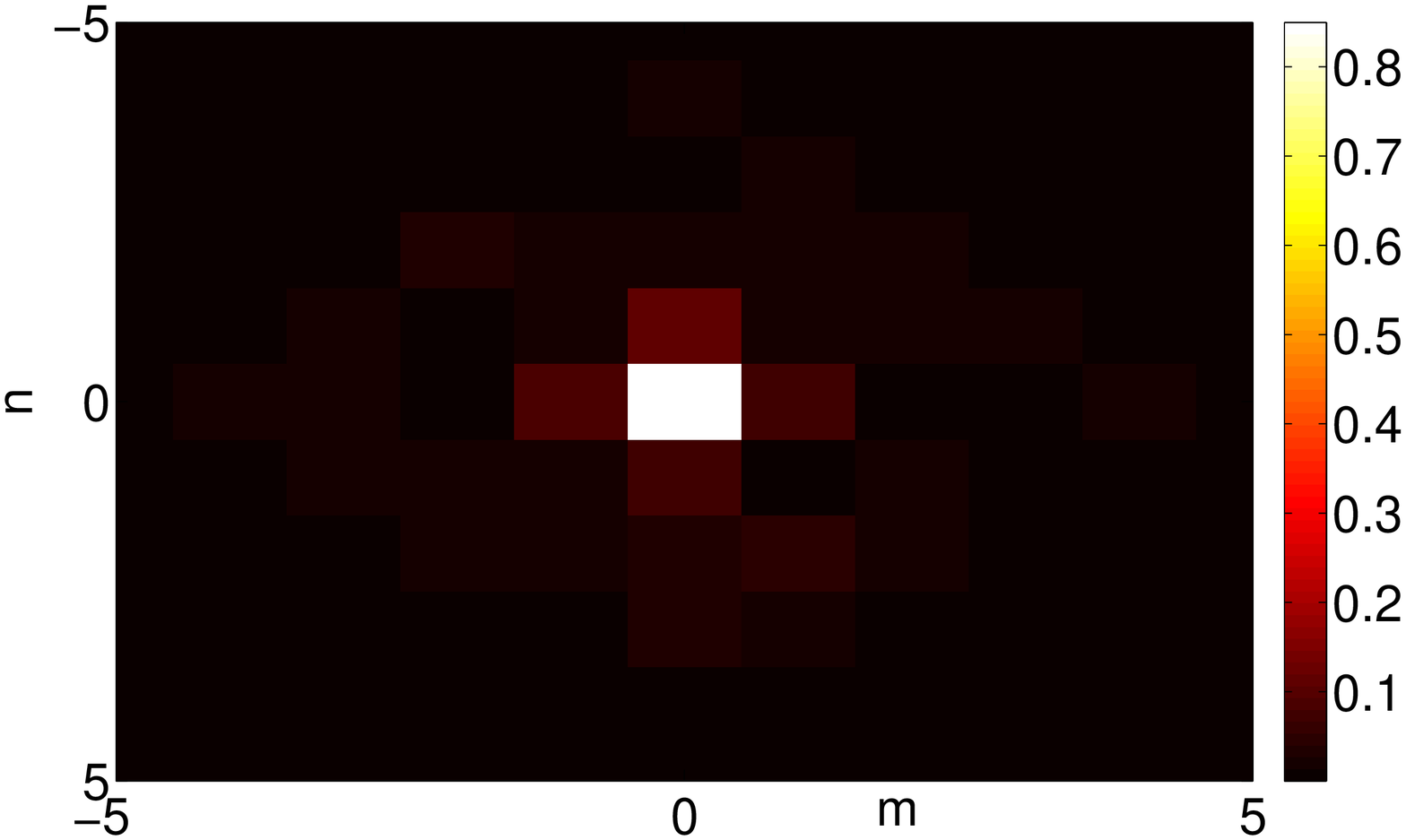} %
\includegraphics[width=0.45\textwidth]{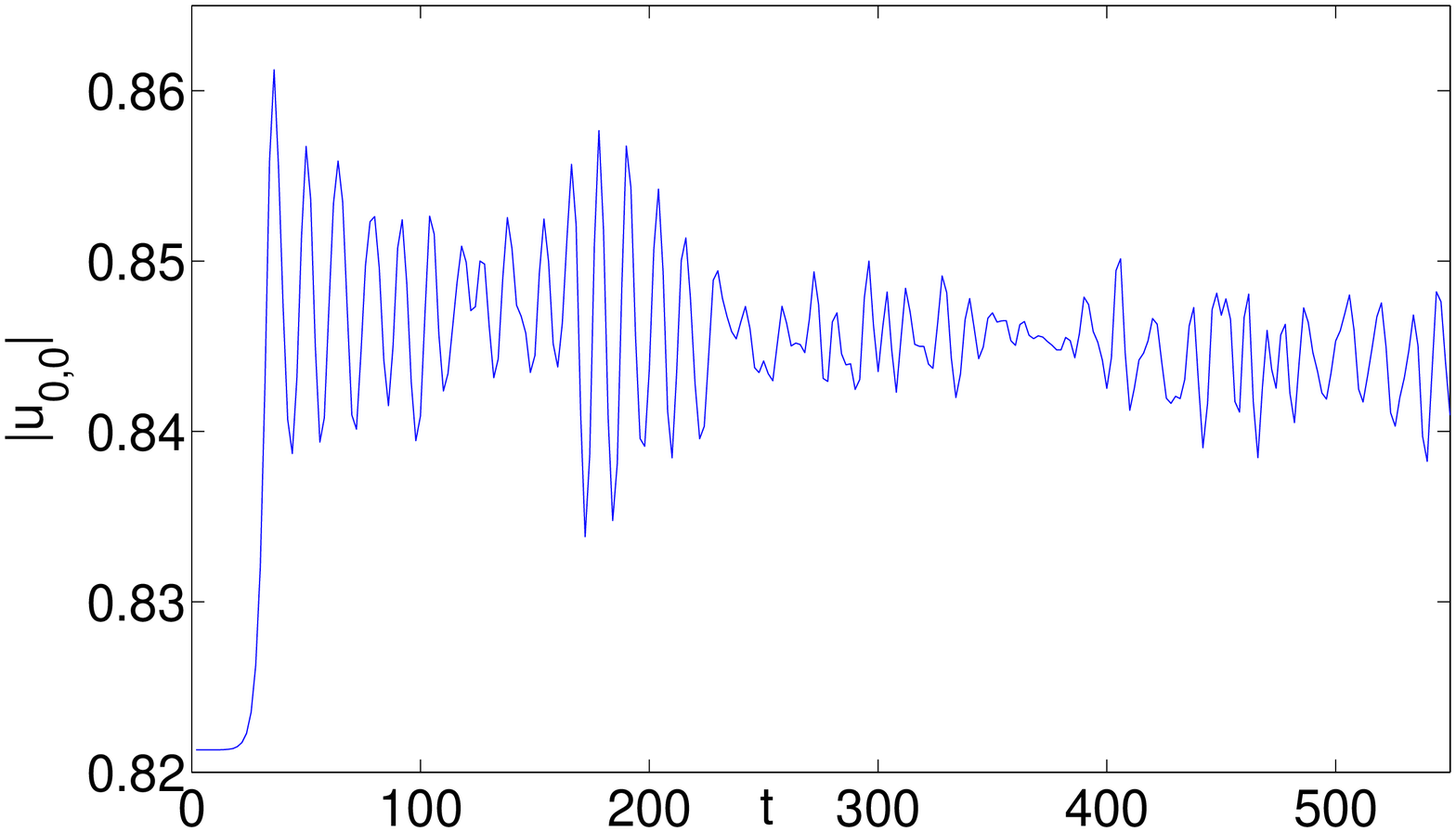}
\par
\includegraphics[width=0.45\textwidth]{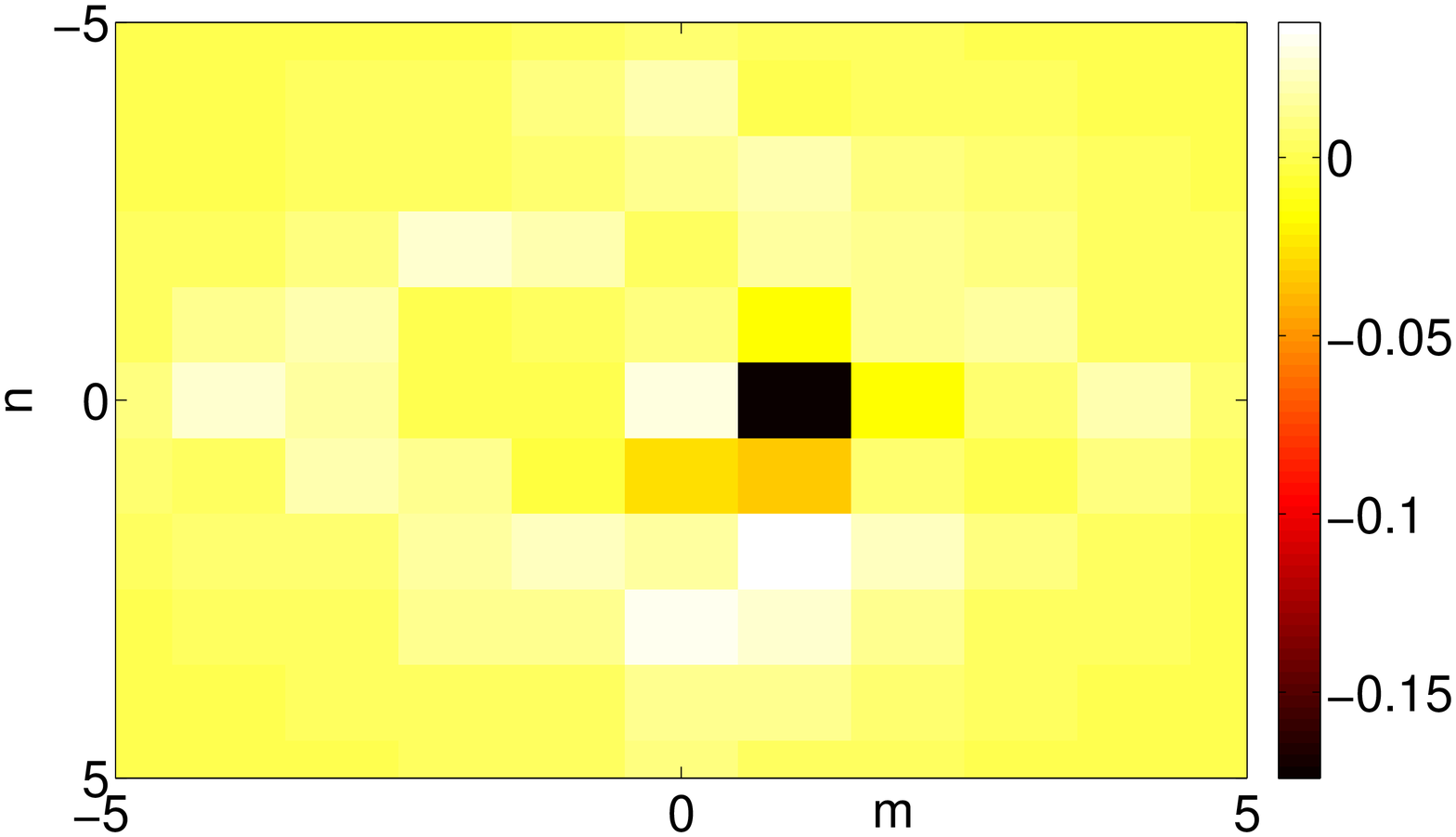} %
\includegraphics[width=0.45\textwidth]{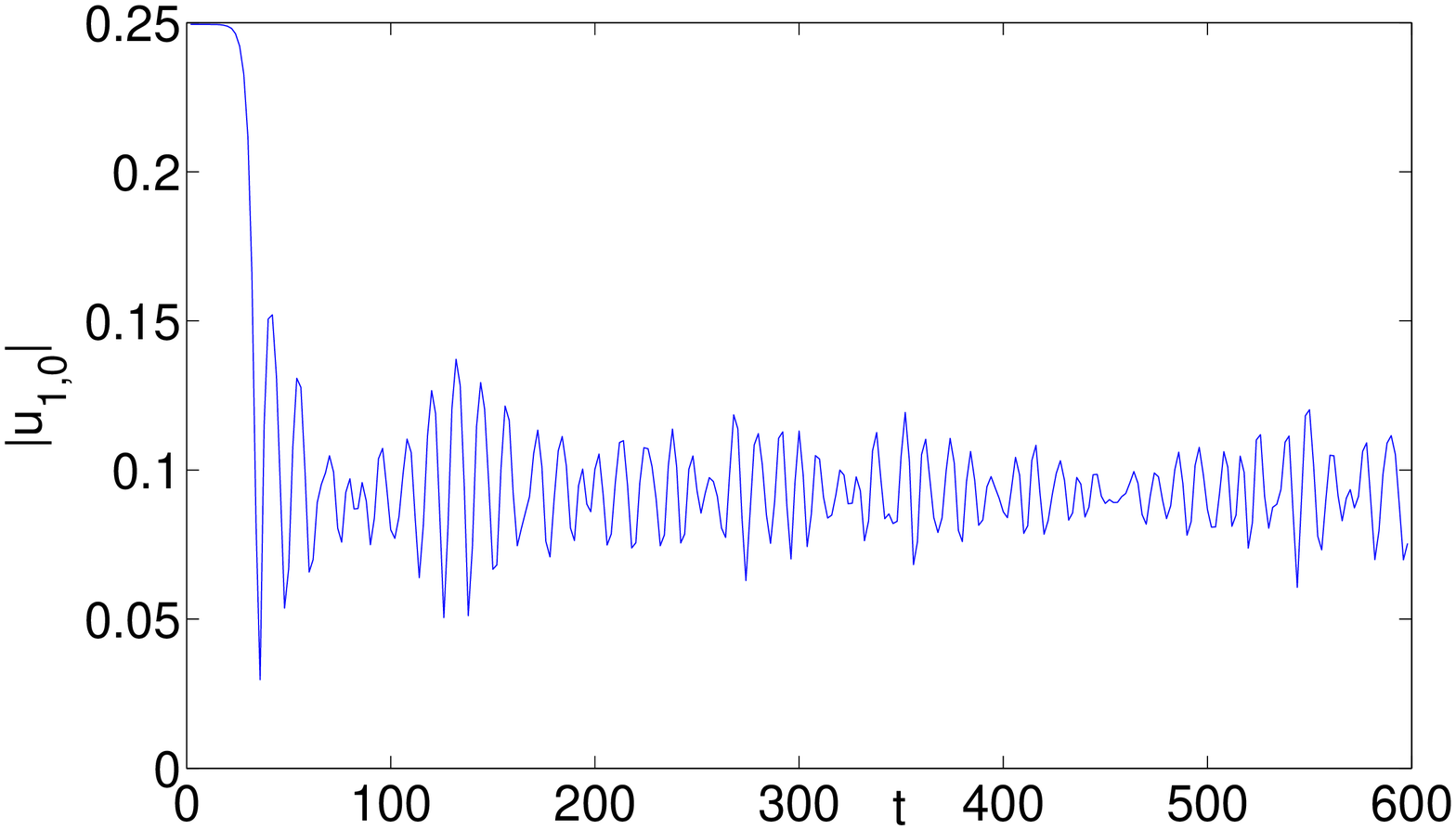}
\end{center}
\caption{(Color Online)
Similar to the previous figure but now for the out-of-phase
twio-site configuration at $\protect\varepsilon =0.07$. The top left panel
shows the result of the evolution at $t=600$, while the bottom left panel
shows its difference from the input, in terms of the absolute value. The
right panels illustrate the evolution at the central (top) and
nearest-neighbor $(1,0)$ (bottom) sites, demonstrating the exponential
nature of the instability.}
\label{new_fig9}
\end{figure}

The evolution of the out-of-phase two-site state for $\varepsilon =0.07$ is
shown in Fig.~\ref{new_fig9}. The top left panel displays the profile
resulting from the dynamics 
at $t=600$, while the bottom left panel illustrates the
spreading of the solution, through its difference from the initial profile.
It is interesting that the structure in the top left panel appears to become
more ``symmetrized" in the course of the evolution, bearing four nearly
symmetric excited sites around the central one. The right panels once again
correspond to the evolution of the central site and one of its neighbors.
Notice that here, as expected, the growth and manifestation of the
instability appear to be exponential, rather than oscillatory.

\begin{figure}[th]
\par
\begin{center}
\includegraphics[width=0.45\textwidth]{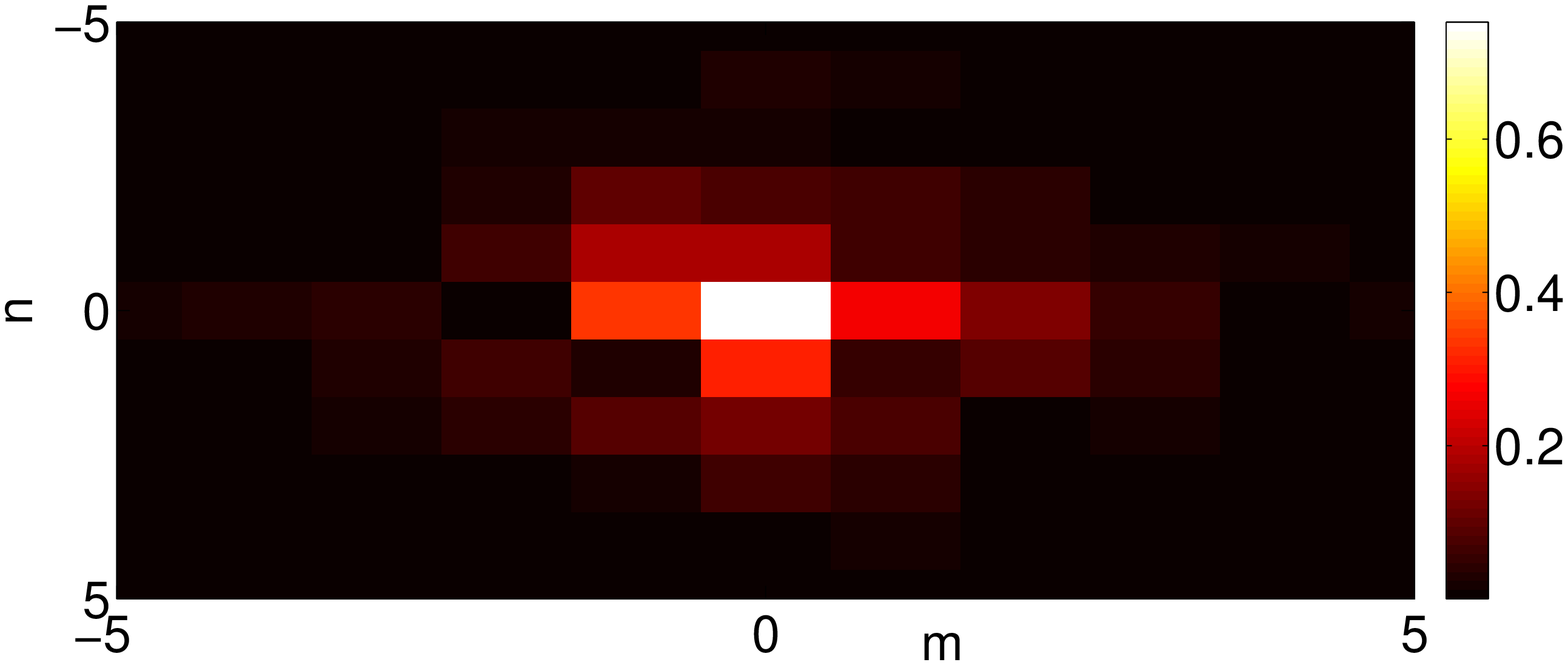} %
\includegraphics[width=0.45\textwidth]{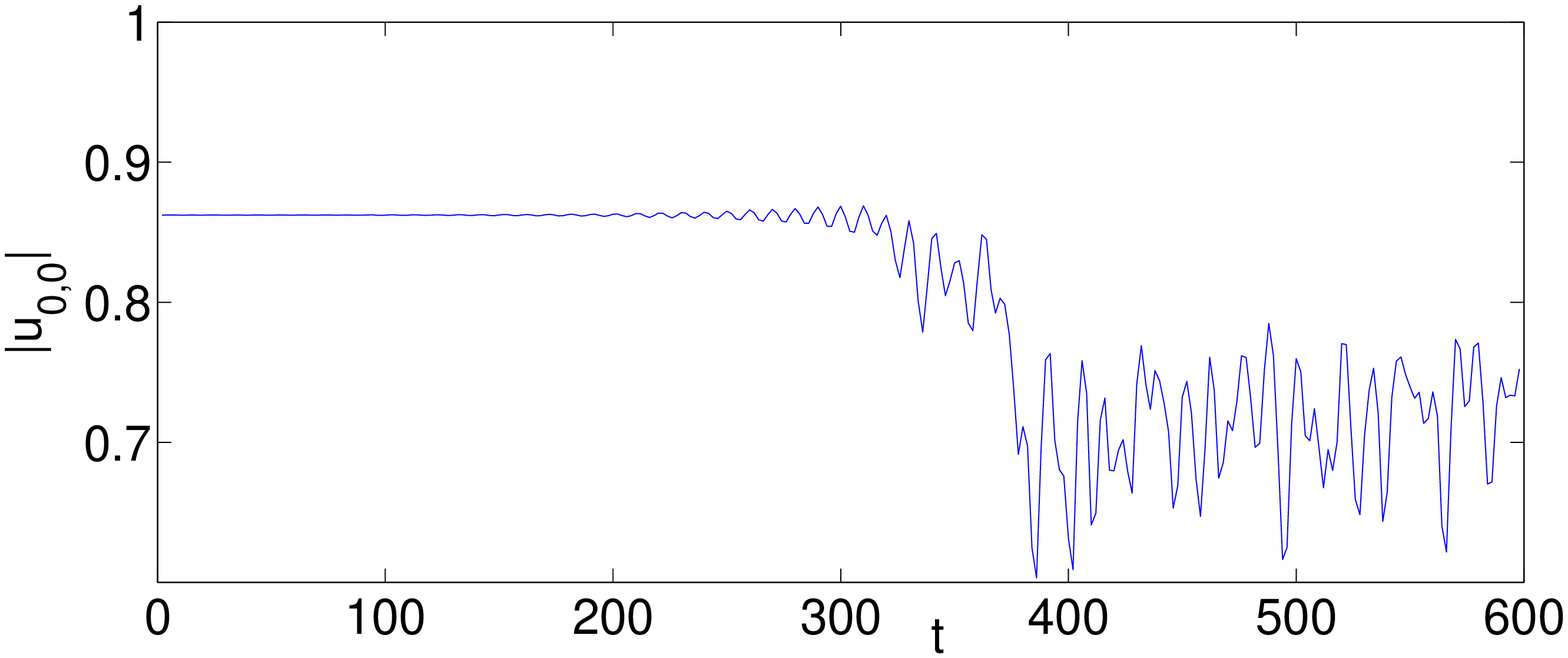}
\par
\includegraphics[width=0.45\textwidth]{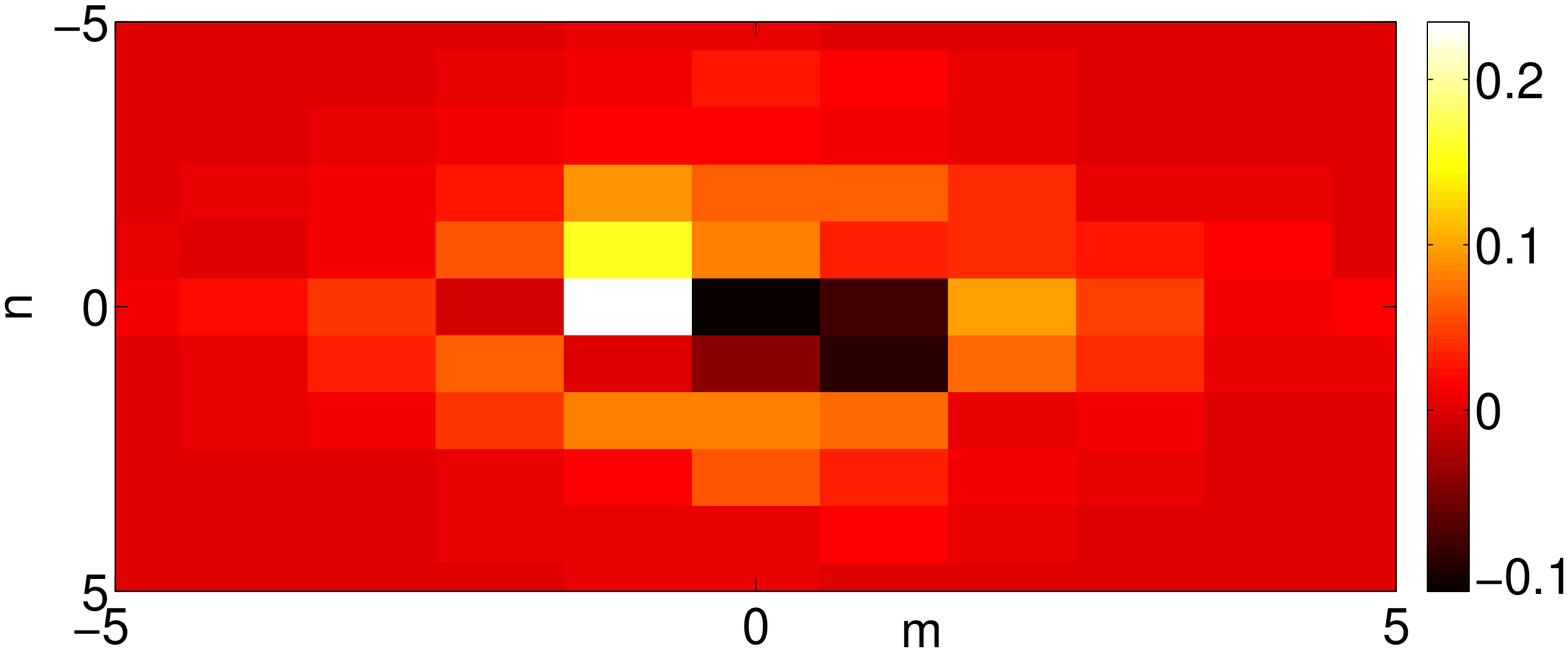} %
\includegraphics[width=0.45\textwidth]{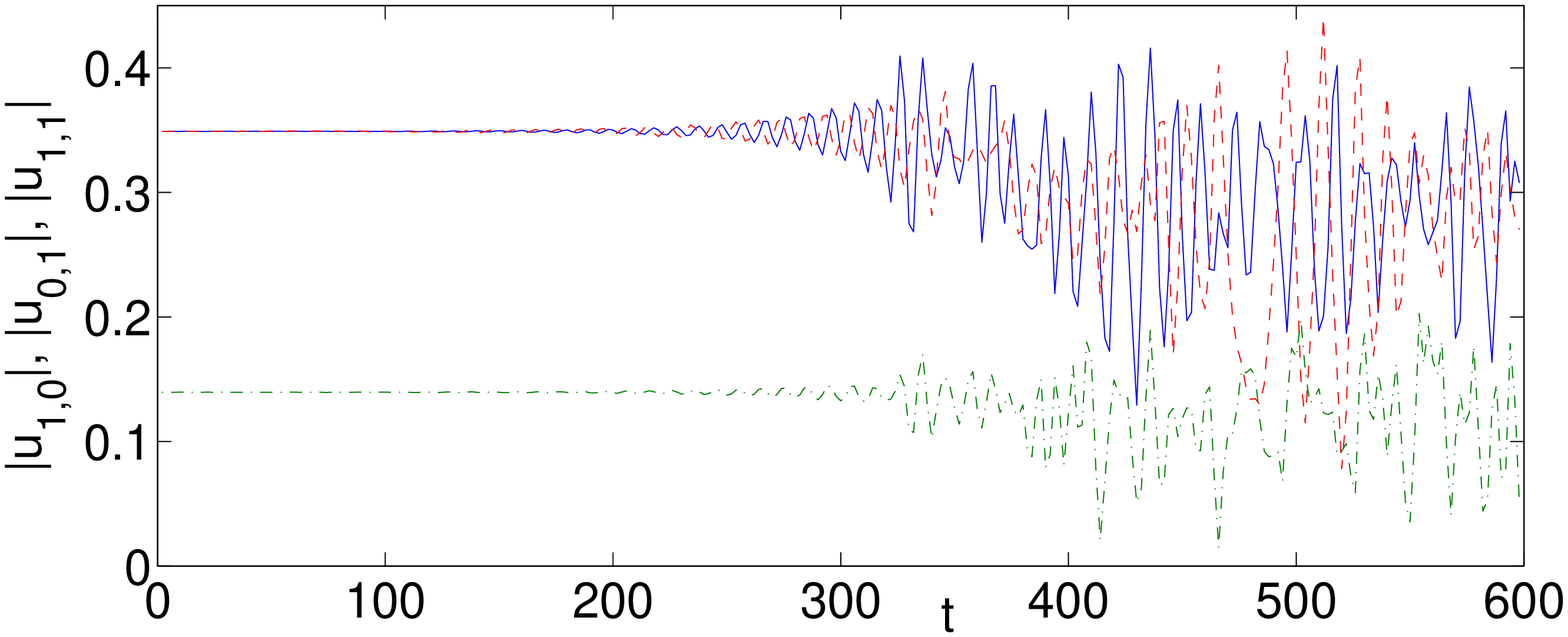}
\end{center}
\caption{(Color Online)
The same as previous figures, but now for the four-site, in-phase
excited state with $\protect\varepsilon =0.075$. Notice the broadening of the
solution (left panels) and the oscillatory manifestation of the instability
(right panels).}
\label{new_fig10}
\end{figure}

\begin{figure}[th]
\par
\begin{center}
\includegraphics[width=0.45\textwidth]{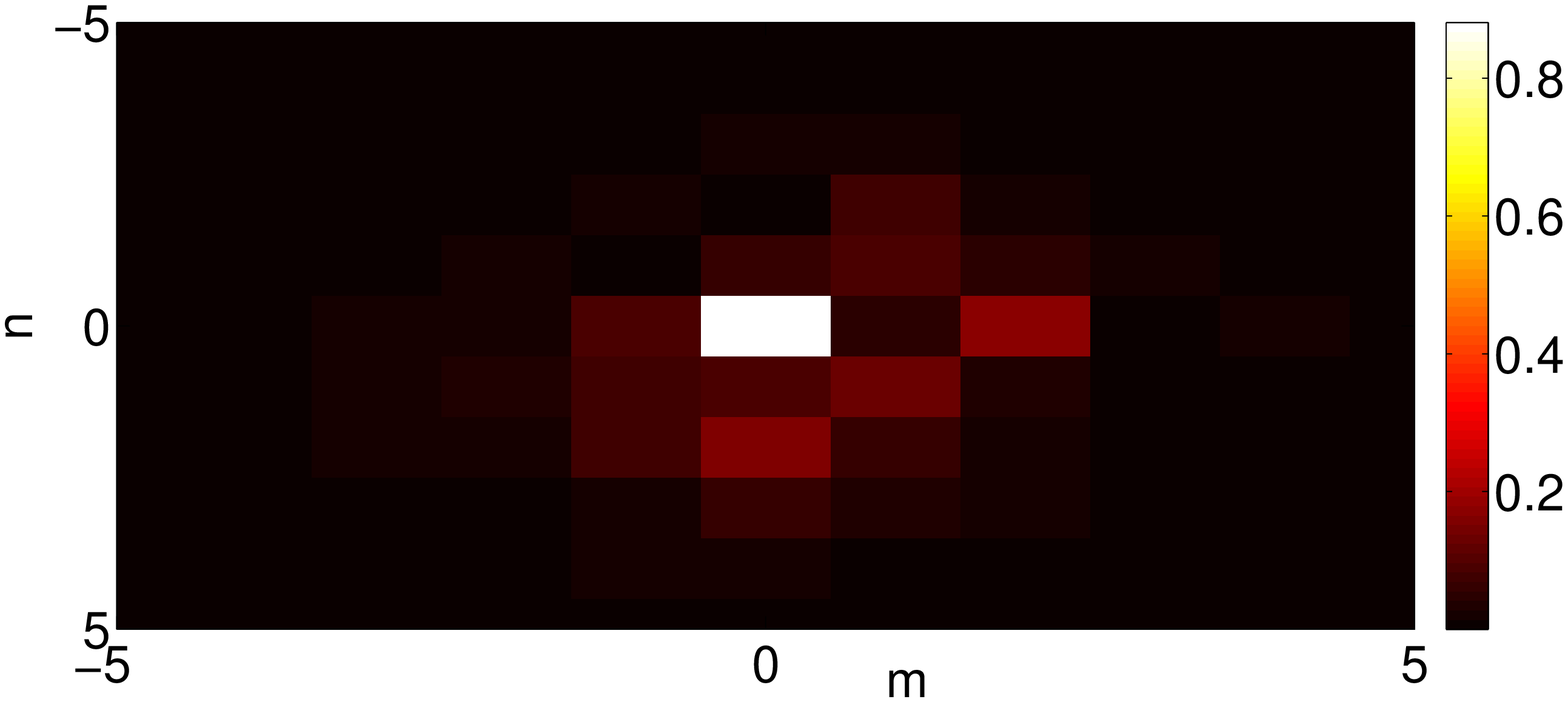} %
\includegraphics[width=0.45\textwidth]{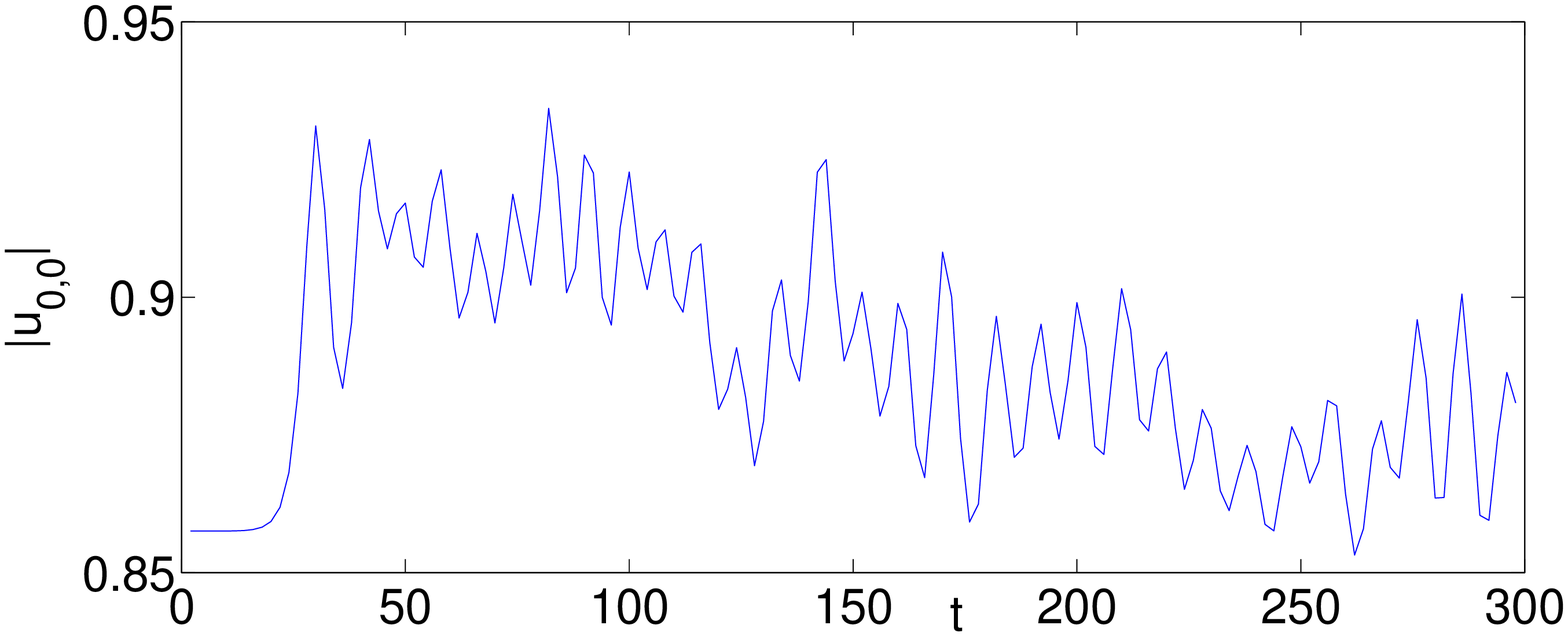}
\par
\includegraphics[width=0.45\textwidth]{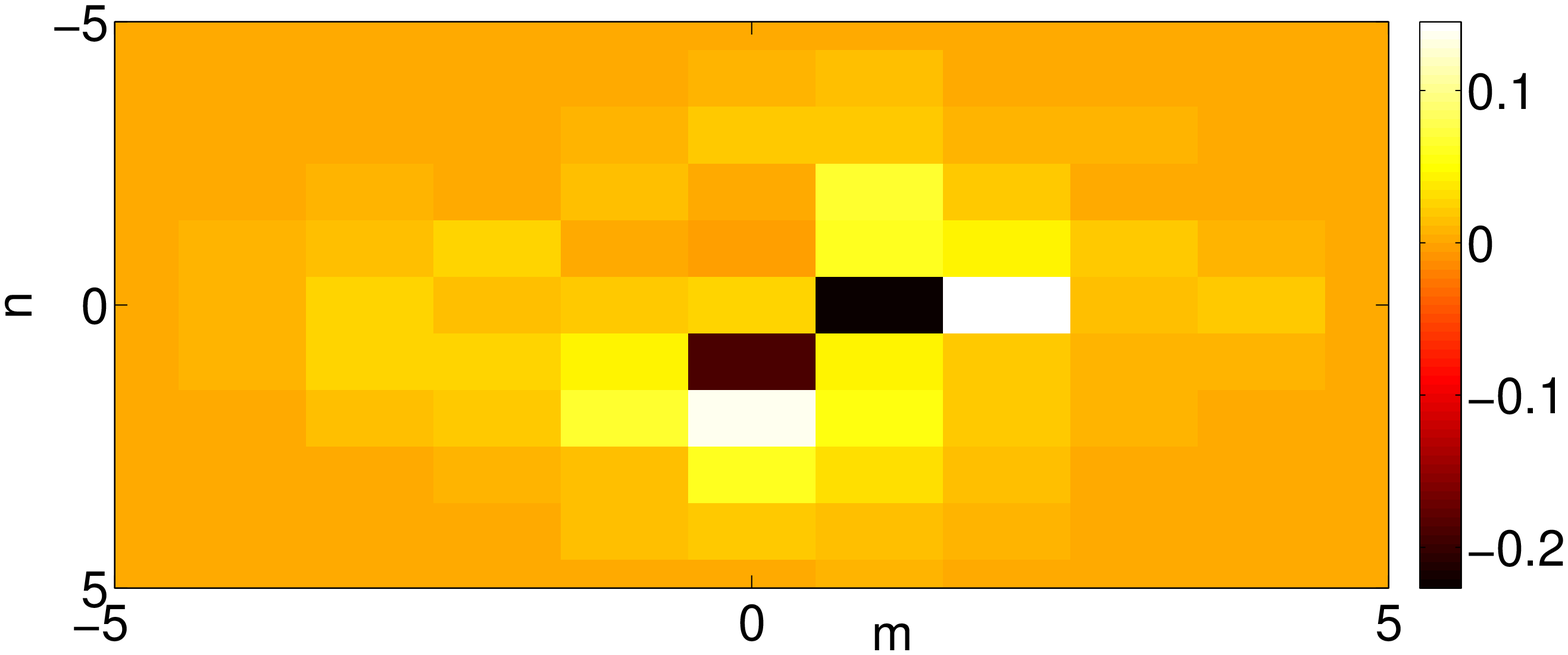} %
\includegraphics[width=0.45\textwidth]{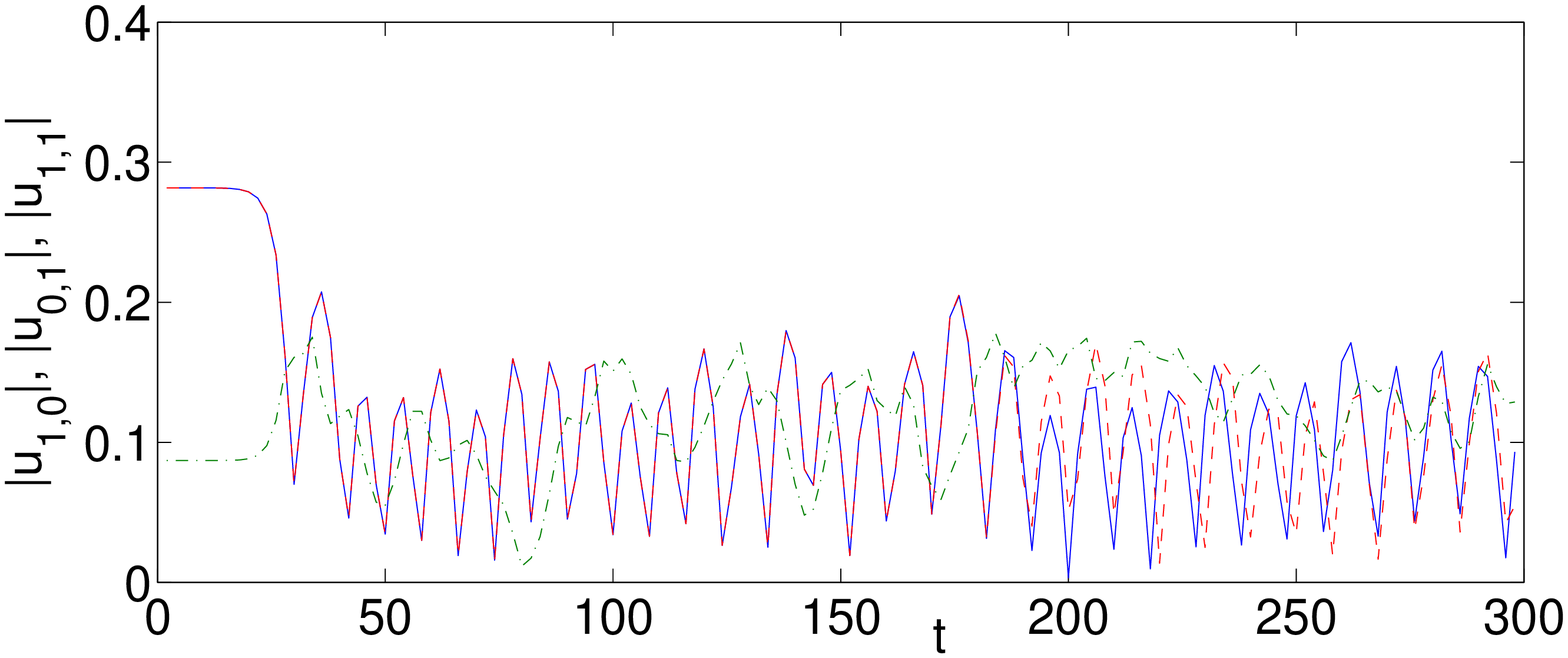}
\end{center}
\caption{(Color Online)
The same as previous figures, but for the exponential instability
dominating the dynamics of an out-of-phase configuration at $\protect%
\varepsilon =0.055$.}
\label{new_fig8}
\end{figure}

Moving to the four-site configurations, we explore the instability of the
in-phase state at $\varepsilon =0.075$ in Fig.~\ref{new_fig10}, and of the
out-of-phase one at $\varepsilon =0.055$ in Fig.~\ref{new_fig8}. The former
state clearly features (see, especially, the right panels) an oscillatory
instability that destroys the configuration, making it broader (see 
the bottom
left panel) and more similar to the configuration with all the sites
excited, which is predicted by the TFA, see Eq. (\ref{TF}) (see 
the top left
panel).

On the other hand, the out-of-phase four-site state with $\varepsilon =0.055$%
, shown in Fig.~\ref{new_fig8}, illustrates an exponential growth, as
illustrated in the right panels of the figure. Here, too, the solution
becomes more extended (see, e.g., the bottom left panel), while its central
amplitude increases, as shown in the top left and top right panels.

\begin{figure}[th]
\par
\begin{center}
\includegraphics[width=0.45\textwidth]{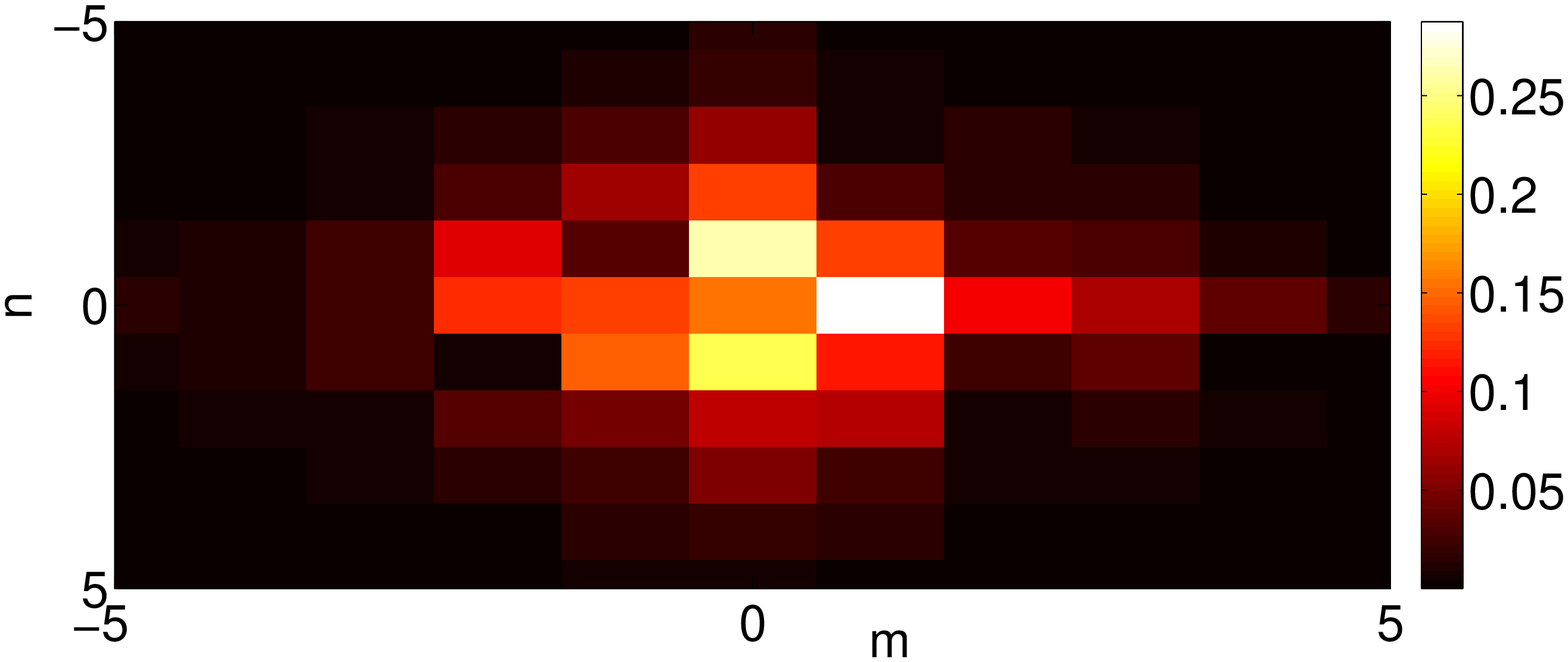} %
\includegraphics[width=0.45\textwidth]{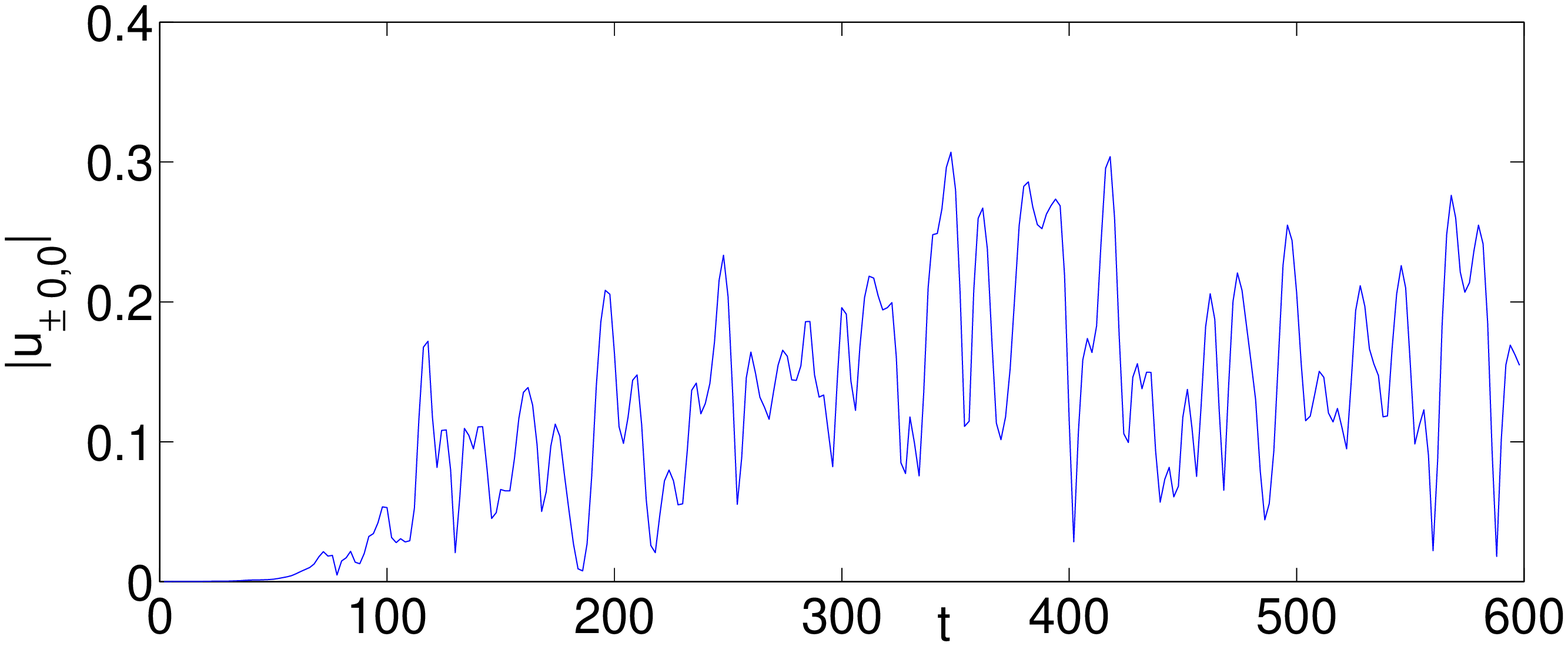}
\par
\includegraphics[width=0.45\textwidth]{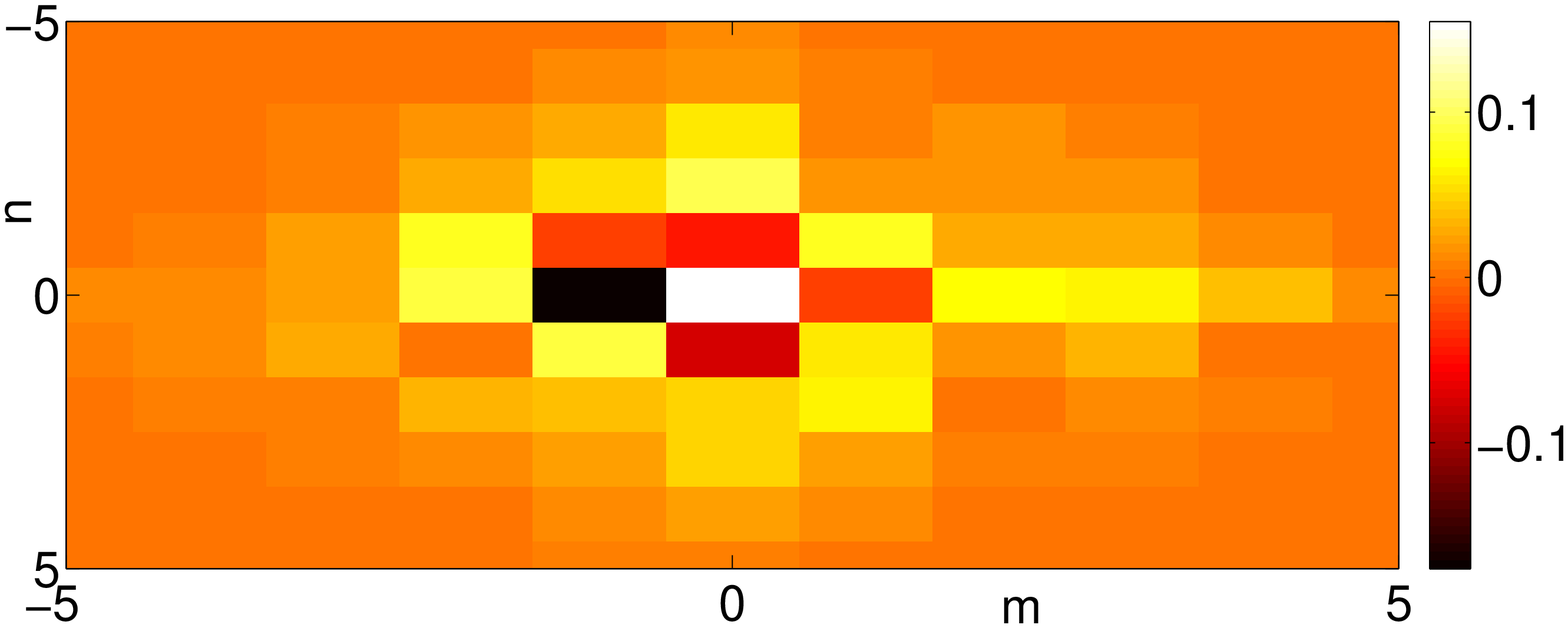} %
\includegraphics[width=0.45\textwidth]{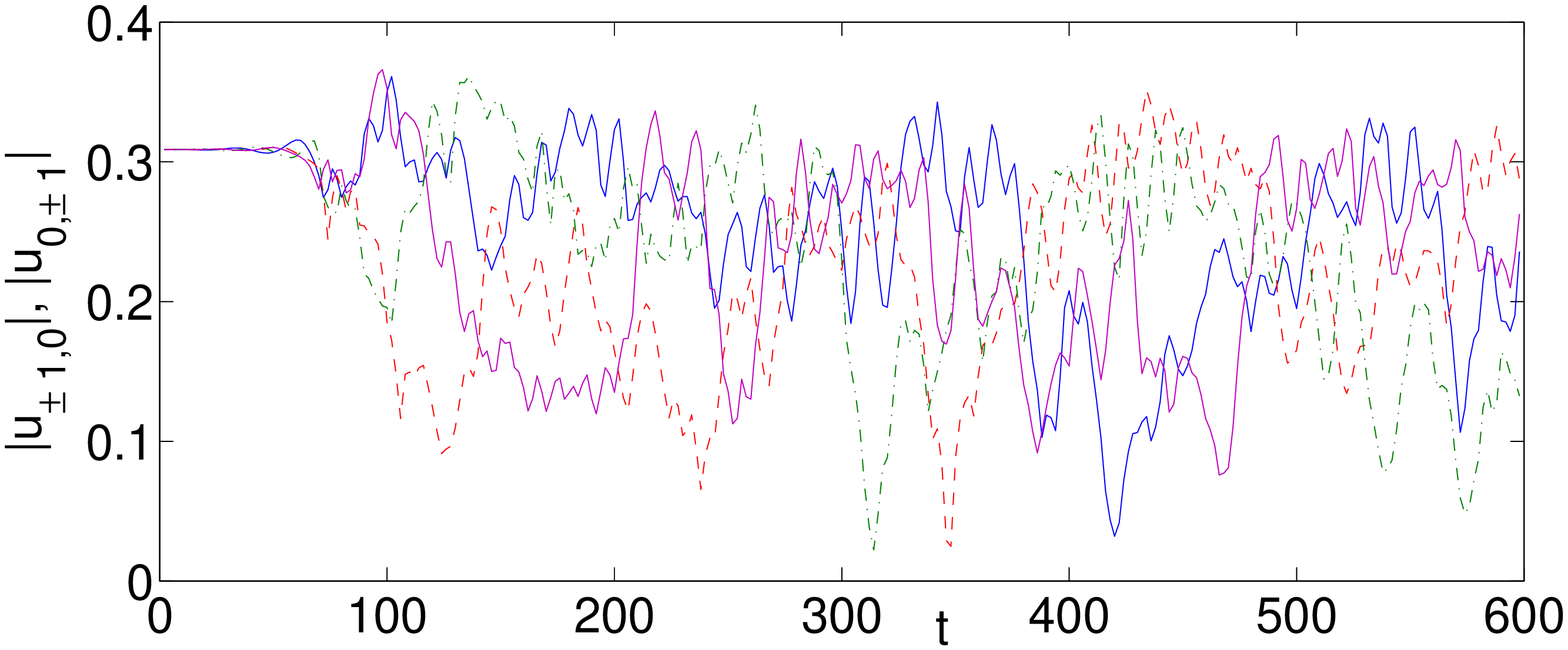}
\end{center}
\caption{(Color Online) 
The same as previous figures, but for the vortex with $\protect%
\varepsilon =0.068$. The top right panel shows populating the originally
empty central site, while the four sites surrounding it no longer bear equal
amplitudes, in the course of the development of the oscillatory instability
(the bottom right panel).}
\label{new_fig11}
\end{figure}

Lastly, the evolution of the vortex configuration from Fig.~\ref{new_fig6}
is displayed in Fig.~\ref{new_fig11}. This configuration too is apparently
destroyed by the oscillatory instability. The latter
 leads to a breaking of the
symmetry of the amplitude pattern built of the four sites which constitute the
vortex (the bottom right panel), as well as to populating the central 
site (the top right panel), which has identically zero amplitude in exact vortex
solutions. The latter effect attests to the destruction of the vorticity, as
is confirmed by the top left panel, which shows that the norm of the
configuration (for the present case of $\varepsilon =0.068$) is spread over
multiple sites surrounding the central ones. This is also made evident by
the difference plot (between absolute values of the initial and the final
configurations) presented in the bottom left panel.

\section{Conclusions and Future Challenges}

In the present work we have explored the existence, stability and dynamics
of different bright solitary waves, as well as vortex configurations, in the
2D defocusing nonlinear Schr{\"{o}}dinger lattice. As in earlier works in
lattices and continua, the fundamental modification of the present setup
enabling the existence of such states is the introduction of the spatially
modulated nonlinearity profile, with the local strength growing from the
center to the periphery faster than the squared distance. As a result,
single-site, two-site (in- or out-of-phase), and four-site (in-, out- or
with mixed-phase) configurations have been systematically constructed near the
AC (anti-continuum) limit. A significant advantage of this construction is
not only its full controllability in this limit, but also the ability to
analyze the linear stability of the configurations. Going beyond these
simplest few-site constructions, we have also explored a vortex cross, as
well as the ``extended" solution, in which all the sites of the lattice are
excited. In fact, solely this last solution was previously 
identified in the 1D
version of the discrete system \cite{malom10}. Here, 
this waveform was found to be
the most robust one, being stable in the entire parametric 
region that was considered. All solutions with out-of-phase structures feature
instabilities accounted for by real eigenvalue pairs, while in-phase 
few-site states are subject to oscillatory instabilities, caused by the
collision of imaginary eigenvalue pairs with the continuous spectrum or
eigenvalue pairs bifurcating from the edge of the continuous-spectrum band.
Monitoring the evolution of the unstable modes, we observed a trend of the
norm to spread over multiple sites surrounding the center, and also an
apparent tendency to rearrange into a structure reminiscent of the
stable extended solution.

There are numerous questions that merit further investigation in this
nascent field. Gaining a more systematic understanding, possibly through
analytical considerations, of vortex crosses (e.g., of their eigenvalue
dependencies) and of vortex squares (in particular, whether they can be
systematically continued from the AC limit to the system with finite
coupling) are relevant directions. Moreover, detailed stability analysis of
the extended bright discrete-soliton configuration should be interesting in
its own right. Extending the present configuration to the 3D setting (cf.
Ref.~\cite{pkf3}) is another challenging issue. On the other hand,
challenging yet interesting too 
should be the extension of the study of quantum solitons in the
Bose-Hubbard counterpart of the present setting from the 1D setting \cite%
{Padova} to 2D. Some of these topics are currently under study and results
will be reported elsewhere.

\section*{Acknowledgements.} The work of D.J.F. was partially supported by the 
Special Account for Research Grants of the University of Athens.
P.G.K. gratefully acknowledges the support of 
NSF-DMS-1312856, as well as from
the US-AFOSR under grant FA950-12-1-0332,  
and the ERC under FP7, Marie
Curie Actions, People, International Research Staff
Exchange Scheme (IRSES-605096). 
P.G.K. and B.A.M. gratefully acknowledge the support of the
BSF under grant 2010239.
This work was supported in part
by the U.S. Department of Energy.

\end{document}